# An exploratory study of the suitability of UML-based aspect modeling techniques with respect to their integration into Model-Driven Engineering context

*Abid Mehmood[1] and Dayang N.A. Jawawi*
*Department of Software Engineering, Faculty of Computing*
*Universiti Teknologi Malaysia (UTM), 81310 Skudai, Johor, Malaysia.*

**Abstract**

**Context:** The integration of aspect oriented modeling approaches with model-driven engineering process achieved through their direct transformation to aspect-oriented code is expected to enhance the software development from many perspectives. However, since no aspect modeling technique has been adopted as the standard while the code generation has to be fully dependent on the input model, it becomes imperative to compare all ubiquitous techniques on the basis of some appropriate criteria.

**Objective:** This study aims to assess existing UML-based aspect-oriented modeling techniques from the perspective of their suitability with regards to integration into model-driven engineering process through aspect-oriented code generation.

**Method:** We defined an evaluation framework and employed it to evaluate 14 well-published, UML-based aspect-oriented modeling approaches. Further, based on the comparison results, we selected 2 modeling approaches, Reusable Aspect Models and Theme/UML, and proceeded to evaluate them in a detailed way from specific perspectives of design and its mapping to the implementation code.

**Results:** Results of the comparison of 14 approaches show that majority of aspect modeling approaches lack from different perspectives, which results in reducing their use in practice within the context of model-driven engineering. The in-depth comparison of Reusable Aspect Models and Theme/UML reveals some points equally shared by both approaches, and identifies some areas where the former has advantage over the latter.

**Conclusion:** Majority of aspect-oriented modeling approaches works well to handle the basic modeling tasks. However, in the context of their integration into model-driven engineering process, these approaches need to be improved from many perspectives. As regards the second part of our comparison, the Reusable Aspect Models approach may be seen as a preferred approach to handle the task of integration using aspect-oriented code generation.

**Keywords:** aspect-oriented modeling, model-driven engineering, code generation.

## 1. Introduction

Aspect-oriented software development approaches are essentially intended to improve the handling of a specific type of concerns, the crosscutting concerns, which represent the functionality that cuts across the primary modularization of a system. Such concerns usually seem to originate from non-functional requirements such as logging, security, persistence etc. However, it is not uncommon to find situations where even the functional requirements have their behavior spread out over several modules. A good example of this is the behavior related to authentication/authorization of a particular role in a software system. A straightforward representation of this behavior at modeling and implementation levels would most likely cut across all modules wherein updates are performed. Such representation results in problems associated with the phenomenon of scattering (a concern spread out over several modules) and tangling (one module

---

[1] Corresponding author: Tel: +966-17-2415533
Email addresses: mabid4@live.utm.my (Abid Mehmood), dayang@utm.my (Dayang N.A. Jawawi)

representing more than one distinct concern) of behavior. Aspect-oriented techniques provide mechanism to explicitly identify, separate and encapsulate such crosscutting behavior (usually referred to as aspect behavior). Once the aspect behavior is encapsulated separately, composition mechanisms are also provided to control where and when this behavior is to be integrated with non-crosscutting behavior (usually referred to as the base). Therefore, in practice, aspect-oriented techniques are applied during analysis using Early Aspects [1], during design using Aspect-Oriented Modeling (AOM) [2], and using Aspect-Oriented Programming [3] for implementation. A few of the benefits of applying aspect orientation to software development may be found linked with maintainability, extensibility and reusability of the system.

The current study is specifically related to aspect-oriented design approaches (i.e. AOM approaches) and their subsequent integration with Model-Driven Engineering (MDE) process. In its essence, MDE process makes models the primary development artifact and uses them as basis for obtaining an executable system. It emphasizes on subjecting models to a refinement process, through automatic transformations, until a running system is obtained. An integration of AOM approaches with MDE process (i.e. obtaining a final executable system from aspect-oriented models) can be realized through two different ways. First approach works purely at the modeling level and uses a model weaver to integrate the aspect and base models in such a way that a non-aspect-oriented (i.e. object-oriented) model is obtained. Object-oriented code generation approaches are subsequently used to generate code into one of the object-oriented programming languages. Second approach directly transforms the aspect-oriented model into code of an Aspect-Oriented Programming (AOP) language and relies on weaver provided by the target language to deal with aspects. We focus on the second approach to integrating aspect-orientation with MDE process in this paper i.e. on aspect-oriented code generation. The motivation behind this selection is provided in the following section. In this context, while we move towards aspect-oriented code generation, the modeling notations used to support aspect-orientation play a vital role, since the completeness of generated code is directly dependent on comprehensiveness supported by the modeling notation used. Moreover, since no modeling language has been adopted as a standard for aspect-oriented modeling, an evaluation of the entire corpus of existing aspect-oriented modeling approaches for their suitability to the purpose of code generation is expected to provide a basis for aspect-oriented code generation. Therefore, this study is focused on evaluating existing AOM approaches in the context of aspect-oriented code generation. In general, this comparison focuses on investigating the suitability of these approaches to serve the goal of integration of MDE and aspect orientation. After evaluating 14 approaches on basis of generally applicable criteria, we proceed to carrying out an in-depth comparison of two approaches i.e. Reusable Aspect Models (RAM) and Theme/UML. To this purpose, we have taken a common example from aspect-oriented modeling literature and modeled it using both approaches. Further, we have also used the same example to investigate the model-code relationship of these two approaches, by mapping the design models of our example to AspectJ code.

In this following, Section 2 provides the motivation for this study and briefly discusses the related work. In Section 3, we define the strategy used for evaluation of AOM approaches, specifically, by describing the rationale behind selection of approaches and the set of criteria used for comparison. Section 4 presents the results of comparison. Section 5 is dedicated to detailed comparison of Reusable Aspect Models and Theme/UML approaches by means of developing design models of our example system and mapping the models to code. Finally, we conclude the paper in Section 6.

## 2. Motivation and related work

In previous section, we have described two different approaches that can be taken to obtain an executable from aspect models i.e. weaving the aspect model to obtain a non-aspect model followed by generation of non-aspect code and transformation of aspect model into aspect-oriented code. The core idea behind the first approach is to provide composition mechanism so that models can be simulated, tested and debugged prior to execution. Several model weavers have been proposed to achieve this. However, the main drawback in this approach is that it does not support separation of concerns that have once been composed during the weaving process. This means that benefits of clear separation of concerns become unavailable after the composition has been performed. This dilemma of losing clear boundaries of concerns while translating models into implementation further leads to problems related to evolvability, traceability, reusability and understandability of the developed software systems (cf. [4]). In contrast, second approach that proposes transformation of an aspect-oriented model directly into aspect-oriented code is mainly inspired by the benefits resulting from existence of a direct mapping between constructs of design model and the programming language. Moreover, several empirical studies for example [5-8] have reported the potential benefits of using aspect-oriented techniques in software development. Another study i.e. Ref [9] has discovered that approaches that target aspect-oriented programming languages result in compact, smaller, less complex and more modular implementations.

Hence, keeping in view the benefits of integration of aspect-oriented techniques with MDE through aspect-oriented code generation, we consider it worthwhile to evaluate existing AOM approaches with respect to their suitability to serve as input to an aspect-oriented code generation process.

Previously, in [10], we have conducted a systematic mapping study of the related area with the aim to identify and classify existing research in context of aspect-oriented model-driven code generation. The results of study indicated the underdevelopment of the area of aspect orientation in general, and aspect-oriented code generation in particular. Some other work has presented comparison of aspect-oriented modeling approaches pursuing some distinguished goals. In this regard, Wimmer et al. [11] have defined a detailed evaluation framework to evaluate existing AOM approaches with focus on comparability in general. The major distinction of their work from all other surveys on the topic is the breadth and depth of evaluation. Chitchyan et al. [12] have presented an extensive work with the goal of "developing integrated aspect-oriented requirements engineering, architectures, and design approaches". Therefore, they have provided review of all significant works on both aspect-oriented as well as non-aspect-oriented approaches to software development. Similarly, Reina et al. [13] have investigated some AOM approaches with specific goal of evaluating dependency of each approach on particular platform and on specific concerns. Op de beeck et al. [14] have presented a comparison of AOM approaches within the context of product line engineering and with the goal to position AOM approaches within software development life cycle for large-scale system. With regards to aspect-oriented code generation, in [15], we have conducted a comparison of 6 approaches, focusing only on the support or otherwise of features required for full code generation.

As far as the integration of aspect-oriented modeling techniques into an MDE environment is concerned, the specific question of the suitability of various approaches for this integration by means of aspect-oriented code generation has not been explicitly investigated so far.

## 3. Evaluation methodology

In this section, we describe the methodology used for evaluation of selected approaches in the context of this study. First, the rationale behind the selection of approaches and a brief description of the approaches is presented. Then we elaborate the comparison approach.

### 3.1. Selection of approaches

Several proposals for AOM have appeared in literature. These proposals differ in perspectives of their distinguished goals, provided notations, level of concern separation, level of abstraction and level of maturity. However, a vast majority of AOM approaches possess at least one common characteristic, which is their extension from UML. There are only two proposals, to best of our knowledge, that do not extend UML for aspect-oriented modeling i.e. Ref [16] and Ref [17]. Extending UML for aspect-orientation seems quite natural and convincing since aspect-orientation can be seen as an extension to object-orientation for which UML is the standard and most widely used modeling language. Therefore, for this research study, focusing only on approaches that extend UML, we have identified 14 well-published, UML-based AOM approaches namely Stein [18], Ho [19], Aldawud [20], Von [21], Clarke [22-24], Jacobson [25], France [26], Pawlak [27, 28], Cottenier [29, 30], Fuentes [31, 32], Katara [33], Klein [34], J.Klein [35-37], and Whittle [38-40]. In the following, we briefly describe these approaches.

1. The Aspect-Oriented Design Model Notation for AspectJ of Stein et al. [18]

Aspect-Oriented Design Model (AODM) provides a design notation specific to AspectJ. Therefore, it extends UML with the only intention to support AspectJ's concepts at the design level. To exploit the huge resemblance between the core concepts of AspectJ and UML, it provides UML representation for basic constructs of AspectJ, namely join points, pointcuts, introductions, and aspects. Mainly the class, statechart and sequence diagrams are used for structure and behavior modeling. They represent join points using UML links, and apply the concept of adopted links to different diagrams in a way specific to each diagram. Similarly, an advice in AspectJ is viewed as analogous to an operation in UML. Aspects are represented as classes of a special stereotype named <<aspect>>. In principle, AODM has been specified using the UML's standard extension mechanism, but for certain specifications meta-model has also been extended. For example, the UML extend relationship, from which the <<crosscut>> stereotype has been derived originally, can be specified between use cases only.

2. The UMLAUT Framework of Ho et al. [19]

*UML All pUrpose Transformer (UMLAUT)* is an open framework for developing application-specific weavers that can generate detailed aspect-oriented design from a high level design modeled using UML. The UML model elements can be input using various formats such as XMI and Java source. Extensions to UML are done through the UML Profile mechanism. The weaving process is implemented as a model transformation applied to the UML model. Specifically, the weave operation is defined as a transformation rule from an initial model to a final one.

3. The UML Profile of Aldawud et al. [20]

The UML Profile for Aspect-Oriented Software development extends the standard UML package structure to define an AOSD package, which is used to encapsulate all elements defined by the AOSD profile. Crosscutting concerns are modeled using aspects, which are extensions to UML core classes. A new stereotype <<aspect>> is used to model aspects, which are further classified into synchronous or asynchronous aspects. In this profile, synchronous aspects are distinguished from asynchronous ones in that they usually control the behavior of the core classes. The <<crosscut>> stereotype is used to model crosscutting relationships. For behavior modeling, this profile does not dictate any specific behavioral package; however, currently only the use of Collaboration and State machine packages has been outlined for this profile.

4. aSideML Notation of Von [21]

The aSideML's Notation is a meta-model extension of UML to support aspect-oriented concepts. Aspects are defined by parameterizing different model elements, and one or more crosscutting interface is defined to organize join point description and the crosscutting behavior of the aspect. Crosscutting features are defined as an extension to the original features of the class. Specifically, to model structure, a new construct called aspect diagram is introduced which extends features of a UML class diagram. Collaboration and sequence diagrams are extended for modeling behavior of the aspect. The join points are defined by means of an enhanced sequence diagram. Weaving of models is also provided which supports the same set of diagrams and generates woven class diagrams, woven collaboration diagrams and woven sequence diagrams.

5.   Theme/UML of Clarke et al. [22-24]

This approach has basically evolved from work on composition patterns [24, 41], and is considered one of the early approaches to aspect-oriented modeling. In this approach, a new declaratively complete unit named "Theme" is proposed at the design level to represent system concerns, which are essentially collections of structures and behaviors inferred from requirements. A distinction has been made between the "base" themes and the "aspect" themes, where aspect themes refer to crosscutting behavior. In Theme/UML approach, an aspect theme is differentiated from a base theme in the sense that in addition to other behavior, it may define some behavior that is triggered by behavior in some other theme. As far as modeling is process is concerned, first the triggered behavior needs to be identified and captured in the form of templates and then the crosscutting behavior related to those templates is modeled. Later, the base themes which are not affected by the crosscutting themes are modeled using the standard UML design process. A different approach is used to modeling of aspect themes by representing them using a new complete unit of modularization similar to a package in standard UML, with stereotype <<theme>>. This theme may comprise of any of the standard UML diagrams to model different views of the structure and behavior required for a concern to execute. Essentially, the aspect theme design is similar to a standard UML package that contains structural and behavioral diagrams. The only difference is the specification of templates listed inside the theme package notation and a sequence diagram for each of the templates grouping in the theme package. Even though Theme/UML allows any kind of UML diagrams to be used for aspect-theme design, package and class diagrams are currently used for structure modeling, whereas sequence diagrams are used for behavior modeling.

6.   Aspect-Oriented Software Development with Use Cases of Jacobson et al. [25]

Jacobson et al.'s approach is based on the idea of using a development process where concerns are kept separated from requirements specification to the implementation phase. For this purpose, they define use case slices to specify high-level design and then refine this design to obtain detailed design. At detailed design level, they represent structure by means of class diagrams and behavior by means of sequence diagrams. Model weaving is not supported. One distinguishing characteristic of this approach is its support for traceability of models pertaining to a specific concern along different phases of software development.

7.   Aspect-Oriented Architecture Models of France et al. [26]

The Aspect-Oriented Architecture Model approach is based on composing model elements that present a single concept using different views. The model elements that are composed using this approach are needed to be of the same type. Aspects may specify concepts that are not present in a target model.  Templates are used in conjunction with package diagrams, class diagrams, communication diagrams and sequence diagram to represent aspects.  In this respect, this approach is similar to Theme/UML approach described previously. The compositor composition mechanism is used to provide the concern composition. Just like Theme/UML, primary models and aspect models are distinguished, where the latter represent crosscutting behavior. Later, a tool called Kompose [42]has also been developed which uses the composition technique proposed by Aspect-Oriented Architecture Models approach.

8.   The UML Notation for AOSD of Pawlak et al. [27, 28]

This notation is a UML profile based on UML 1.x to model a design using JAC Framework, which is a middleware to support concerns such as persistence, security, fault tolerance etc. in J2EE applications. Currently, the profile does not support behavior modeling, whereas the support for structure modeling is provided by means of class diagrams. <<aspect>> stereotype is used to represent aspects, and they are linked with a target class using <<pointcut>> stereotypes. The association between the operations of base and aspect classes (i.e. the join point) is specified with the help of a proprietary language. This approach is similar to that of Jacobson et al. described previously in that it does not support model weaving.

9. The Motorola WEAVR Approach of Cottenier et al. [29, 30]

The Motorola WEAVR approach has been developed in an industrial setting, specifically in context of telecommunication industry. It uses Specification and Description Language (SDL) to specify models, which has partly been adopted in UML 2.0. The WEAVR approach is thus based on UML 2.0. The approach uses class diagrams and composite structure diagrams to represent structure. State machines, action language of SDL and sequence diagrams are used to model behavior. Individual aspects are represented using <<aspect>> stereotype and a pointcut-advice mechanism is used for composition of aspects and target models. Model execution and code generation are also supported.

10. The Aspect-Oriented Executable Models Notation of Fuentes et al. [31, 32]

Aspect-Oriented Executable UML Models (AOEM) is a UML profile that extends the UML and its action semantics to construct aspect-oriented executable models. In AOEM, an aspect is represented by a UML class stereotyped as <<aspect>>, and comprises special operations to model advices. Specifically, advices are modeled using activity diagrams without input objects and any number of output pins to modify values of intercepted objects. In [43], a dynamic weaving mechanism has also been provided by authors of the AOEM to enhance its models.

11. The Concern Architecture View Approach of Katara et al. [33]

Katara et al. have provided a conceptual model for design of aspects by designing a concern architecture model. The approach can handle the specification of aspects both in symmetric as well as asymmetric manner. The conceptual model has been implemented as a UML Profile. Aspects are defined as augmentations to target model, and the composed model is obtained by mapping a non-aspect model to a new model containing aspect descriptions. The aspects are parametric in nature and thus can be bound or instantiated several times. To make them generic in this way, they are explicitly split into two parts: *required* (which defines the join point*)* and *provided* (which defines the augmentation to original model). To compose an aspect into a target model, a special operation called superimposition is used, which allows an aspect to augment another one.

12. The Behavioral Aspect Weaving Approach of Klein et al. [34]

This approach is based on the concept of scenarios which are basically sequence diagrams or Message Sequence Charts. A pair of scenarios is used to define an aspect, one scenario representing the pointcut, and the other representing the advice. Just like AspectJ, the advice in this behavioral aspect weaving approach can be inserted "around", "before", or "after" a join point. In order to weave an aspect model into a target model, first a generic detection strategy is used which identifies all the join points in the target model, then a generic composition mechanism is applied to compose advice model with the target model at the identified join points.

13. Reusable Aspect Models of Klein et al. [35-37]

Reusable Aspect Models (RAM) is a multi-view modeling approach that combines existing aspect-oriented approaches to model class, sequence and state diagrams into a single approach. Multi-view modeling, in its essence, provides means to describing a system from multiple points of view, using different modeling notations, and thus allowing the use of the most appropriate modeling notation to describe facets different views of a system. RAM is different from all other AOM approaches in a sense that it views aspects as concerns that are reused many times in an application or across several applications. Therefore, this approach models any functionality that is reusable by means of an aspect. Hence, different views (i.e. structure, message, and state views) of a reusable concern are encapsulated in the form of an aspect model which is essentially a special UML package. This aspect model comprises of three different compartments representing the structural view, state view and message view. These views are expressed using a UML class diagram, state diagram and sequence diagrams respectively.

14. MATA notation of Whittle et al. [38-40]

Modeling Aspects using a Transformation Approach (MATA) is a graph transformation based approach to modeling and composing aspects. Both the base and aspect models are represented in the form of well-formed graphs. Since the idea of using graph rules is broadly applicable, MATA is seen as an approach which can be extended to any modeling diagrams, and even to other modeling languages. The only condition in this regard is that the modeling language to be represented using MATA must have a well-defined meta-model. UML meta-model can be represented in the form of a graph by making each meta-class a node in the type graph, and making each meta-association an edge in the type graph. In this way, any UML model can be represented as an instance of this type graph. Aspects are defined using graph transformation rules, where the left-hand-side (LHS) of a transformation rule is a pattern that defines the pointcuts which are to be matched, whereas the right-hand-side (RHS) defines new elements to be added (or removed) at these pointcuts. MATA provides a convenient way of writing the graph rules by proposing that the rule be given on one diagram only, rather than writing graph rules using both LHS and RHS, since this needs repetition of unchanged elements on both sides of the rule. To this purpose, it defines three new stereotypes namely: (1) <<create>> to specify that a new element must be created by the graph rule, (2) <<delete>> to identify deletion of an element by a graph rule, and (3) <<context>> to avoid effect of (1) and (2) on elements.

### 3.2. Comparison approach

A summary and categorization of our comparison criteria is shown in Figure 1. It has to be noted here that the actual comparison criteria employed in this study is, in essence, inspired by all existing surveys on AOM approaches discussed previously in Section 2. However, we have adapted it, in some cases semantically, to fulfill needs of this study in the specific context of aspect-oriented code generation. Moreover, we have extended the comparison to all existing well-published AOM approaches.

In general, the selected criteria refer to revealing relevant information from two broader perspectives. First, we try to analyze each studied approach in terms of its overall suitability to be used for the purpose of code generation, and in turn, for the long-term goal of automatically obtaining useful code in practical scenarios. The analysis from this perspective seems particularly relevant since the modeling approach, in general, is expected to be robust and mature enough to be integrated into broader contexts of model-driven engineering and automatic software systems development. Second, the devised criteria are representative of the analysis leading to defining the likelihood of each approach to effectively work for a practical code generation system.

We base our comparison given in the following sections on the set of criteria described below:

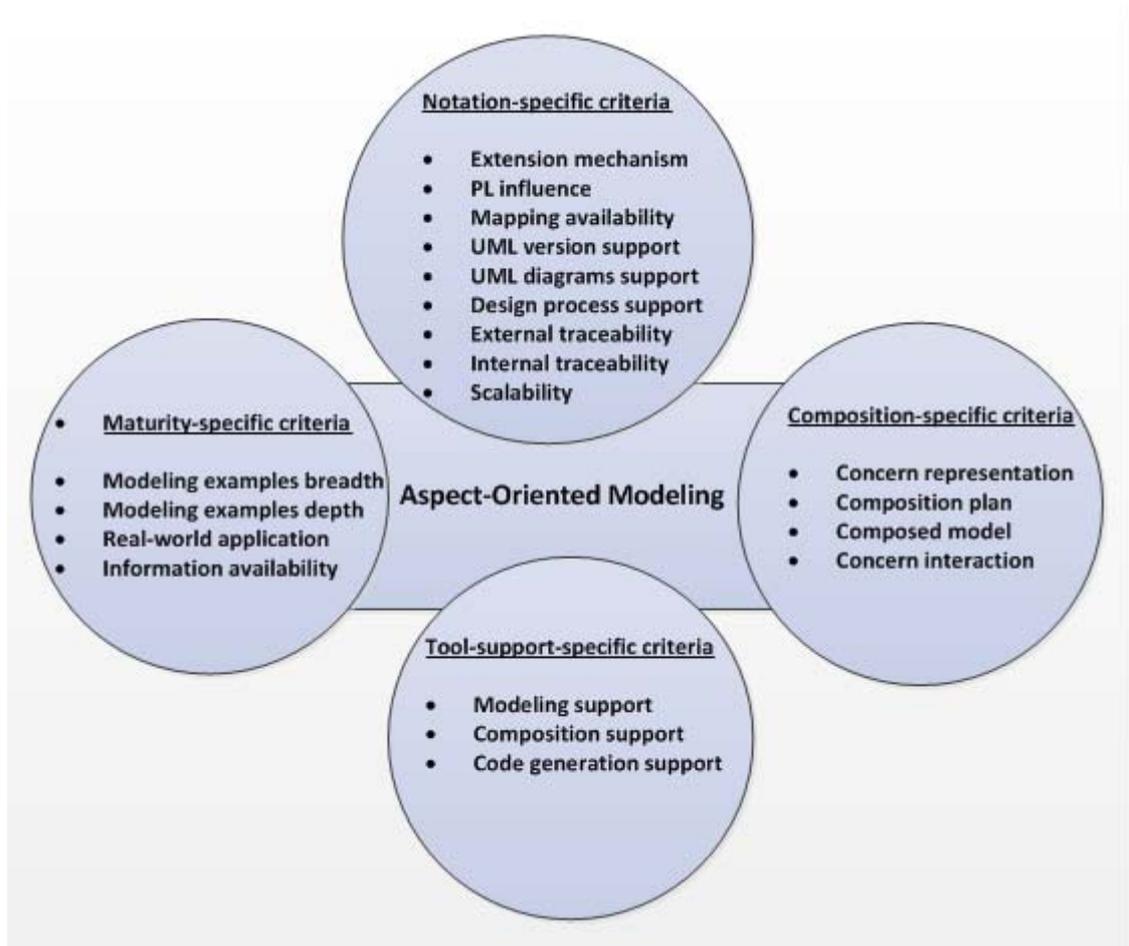

Figure 1: Comparison criteria

### 3.2.1. Notation-specific criteria

*Extension mechanism:* This criterion is used to explore if the AOM approach extends UML by extending meta-model of UML (heavyweight extension mechanism), or by defining a UML profile (lightweight extension mechanism). In fact, both approaches to extending UML have their pros and cons. In principle, however, it has been shown previously (cf.[13, 44] that extending the meta-model is preferable in case if aspect-oriented concepts were well-defined, stable and their likelihood to transfer to or composition with other domains was little. On the other hand, since UML profiles are nothing but a new dialect of UML to support specific domains [45], they are considered ideal in situations where the aspect domain is subject to change and evolution, and it is likely that it will transfer to, or be composed with other domains. Yet another important factor that relates to the distinction between meta-model extension and profile mechanisms is the tool support for UML extensions. Since UML-based tools are built on the meta-model, any changes in meta-model essentially require corresponding adjustments to UML tools as well. However, since UML profiles use standard extension mechanisms and disallow changes in the meta-model, no changes are needed for existing tool support. We have defined this criterion with intention to highlighting related information in the studied approach and to make its correlation with code generation approaches.

*Programming language influence:* As noted by Reina et al. [13], constructs and their respective handling in AOM languages are often inspired by concepts expressed in a specific aspect-oriented programming language. This criterion identifies the specific AOP language that influences the given AOM approach. An investigation of the use of language-specific constructs is fundamental while considering code generation to a specific target language.

*Mapping availability:* In addition to the influence of a programming language on the development of an AOM approach, we also investigate if there is some work related to mapping of the models in given AOM approach to certain aspect-oriented language. This criterion actually identifies the programming language for which mapping has been proposed, only if a mapping definition is provided as a proof, or some appropriate examples are given. This criterion is understandably relevant in context of code generation. Therefore, we will consider it in detail while discussing the likelihood perspective.

*UML version support:* This criterion refers to the UML version supported by the surveyed AOM approach. Main purpose of this criterion is to identify the approaches that do not support the current major enhancements of UML. This criterion is essentially related to looking into suitability of the approach (the first perspective).

*UML diagrams support:* This criterion identifies and lists all different types of UML structural and/or behavioral diagrams extended by the given AOM approach, in order to support aspect-orientation. This criterion is related to both perspectives of our study. An AOM approach cannot be called effective unless it provides a mechanism for a comprehensive modeling of the system, through supporting the effective diagrams. Furthermore, this criterion is much more relevant for code generation as the amount of code generation may directly be influenced by the diagram(s) used as input to code generation approach.

*Design process support:* A design process is relevant in our research study as the completeness of design is an essential requirement for generating complete and fully executable code. This criterion evaluates the surveyed approach in terms of its support for the design process. Specifically, it investigates if this support has been provided in an explicit way, by supporting the detailed modeling, or only implicit support is available, for example in terms of guidelines.

*External traceability:* The external traceability criterion refers to the relationship between two models at different phases in full software development life cycle, where one model is refinement of another. In particular, this criterion investigates if the surveyed AOM approach provides traceability of models over phases such as requirements specification, design, and implementation. In case this traceability is supported, it lists the combination of phases over which the support is provided. For example, an external traceability support from requirements specification to design and to the implementation level will be specified as R → D → I. Support for external traceability is important in the context of our study as the main rationale behind the use of aspect technology is to maintain traceability through different phases of development life cycle.

*Internal traceability:* Unlike external traceability which focuses on different phases of development life cycle, internal traceability is concerned with traceability between two models belonging to one phase. Specifically, this criterion will explore whether mechanisms or guidelines are provided for applications of the approach in which abstract models are refined (i.e. traced) into detailed models, for example, refinement of an abstract design model into a detailed design model. The support for such traceability is very important while studying prospects along the transformation of models into other models and eventually into code.

*Scalability:* Scalability of an AOM approach can be defined as its ability to handle both small as well as large modeling scenarios. The scalability criterion here refers to investigating if the scalability has been proven in real-world projects or at least by examples that go beyond composition of two or three concerns. The rationale behind investigating scalability is that an AOM approach that lacks in scalability cannot obviously be used to generate code which is of industrial standard.

### 3.2.2. Composition-specific criteria

*Concern representation:* This criterion refers to a concern's representation in the AOM approach in terms of a UML meta-class or a stereotype, and identifies the actual notational element used. For example, some AOM approaches use a stereotype <<aspect>> to represent an aspect distinguished from the meta-class Class. The representation of concern is relevant from both perspectives investigated in this study i.e. the suitability as well as the likelihood for extension to code generation. This representation actually encompasses certain characteristics on basis of which the *element symmetry*(cf.[4]), *composition symmetry* (cf. [14]) and *relationship symmetry* (cf. [4]) are determined. Specifically, the element symmetry refers to defining whether there exists some distinction between the concern modules of different structures (i.e. those representing crosscutting concerns versus others), or all concerns are treated equally with modules of identical structure. The composition symmetry is related to defining which concern modules can be composed with each other. Therefore, asymmetric concern composition implies that only "aspects" are allowed to be woven into "base". On the other hand, symmetric composition does not make any distinction between the base and aspects, and allows all types of concerns to be composed with each other. Similarly, relationship symmetry (sometimes referred to as rule symmetry (cf. [11]), determines the placement of composition rules for concerns. The approaches in which composition rules are given within one of the modules that are to be composed are asymmetric in this perspective. A common example of this is the AspectJ language which encapsulates composition rules within an aspect in the form of pointcut-advice combinations. The approaches which are symmetric in this regard define the composition rules in neither of the modules. Therefore, from modeling perspective, the concern representation is actually what will influence its composition, whereas from code generation perspective, it is the abstraction that will somehow need to be transformed into some abstraction at the code level.

*Composition plan:* This criterion evaluates if the support for integration of concerns at the modeling level is provided or not. Lack of this support at the modeling level would mean that composition is deferred to later phases such as implementation (with the help of an aspect-oriented programming language). In case, a composition plan is available, it is investigated whether it is dynamic or static. Dynamic composition refers to support for integration of concerns at the model run-time i.e. during the execution of the model. Composition of models is particularly important for the purpose of simulation and validation of models. In this way, it seems relevant to both perspectives of this study.

*Composed model:* This criterion evaluates the output of concern composition in cases where static composition plan is applied. In particular, it investigates if the output composed model is represented using standard UML or some extensions to UML. Just like composition plan, this criterion will also provide insight into modeling notation in terms of its suitability, even though it is not much relevant from code generation perspective. This is because our intention is to explore aspect-oriented code generation, whereas a woven and standard UML model will eventually be consisting of only object-oriented constructs.

*Concern interaction:* This criterion is specific to interactions between various concerns in a model. In particular, it evaluates if interactions between concerns can be modeled, and if provided, it identifies the notations used to represent this interaction, for example, a meta-class or stereotype. From code generation perspective, the transformation approach and processes thereof will have to provide details of translation of these interactions at the code level.

### 3.2.3. Maturity-specific criteria

The criteria in this section are intended to evaluate the maturity of the studied approaches. In this section, maturity is specifically measured in terms of existing information proving the applicability and pragmatism of an approach.

*Modeling examples breadth:* This criterion is intended to evaluate maturity of given AOM approach by investigating the number of examples modeled using the approach.

*Modeling examples depth:* Apart from investigating the breadth of modeling examples, we measure the depth of examples as well; by looking at how many different concerns have been integrated within the modeling examples. Therefore, this criterion investigates the maximum number of concerns integrated in one example.

*Real-world application:* This criterion evaluates the successful application of surveyed AOM approach in the design of a real-world project. This measure is expected to show the applicability of approach, which as a result, indicates the level of maturity.

*Information availability:* This criterion refers to the available amount of papers, manuals, books and other relevant material associated with the given AOM approach. Specifically, this criterion provides the number of different resources of information on the modeling approach.

### 3.2.4. Tool-support-specific criteria

*Modeling support:* This criterion is related to evaluation of different means to use the surveyed AOM approach to design. Moreover, it also refers to investigation of support for syntactic and semantic validation of the design.

*Composition support:* It refers to the tool support for composition of concerns at the modeling level. This support, if provided, may specifically allow simulation, validation and debugging of models.

*Code generation support:* In line with the concepts of model-driven engineering, AOM notations occasionally address mapping of design models to implementation code, and subsequently support the code generation from models. This criterion evaluates if code generation has been addressed, at least in principle, by the surveyed AOM approach.

## 4. Evaluation results and discussion

Table 1- Lack of *tool-support for composition:* A majority of approaches has provided tool-support for modeling only which is largely due to their use of UML profile mechanism. On one hand, the support for composition of models is an essential requirement for approaches which handle details of composition at modeling level only and generate object-oriented code, since without this support it would not be possible to obtain a composed view of the system. On the other hand, it is desirable for other approaches as well that defer composition to implementation level, since in this case the same can be used for activities such as simulation and validation of model. As shown in Table 4, half of the approaches have not discussed any mechanism for composition of models.

Table 4 summarize the results of comparison of various AOM approaches from notation, composition, maturity, and tool-support perspectives, respectively. While certain observations are evident, some interesting findings are summarized and their results are illustrated in the following.

### 4.1. Notation

Table 1 summarizes the evaluation with regard to the *Notation* category.

*Balance in extension mechanisms and UML versions:* Currently, to a certain extent there is a balance between meta-model extensions and UML profiles. As described previously, UML profiles have some benefits with regard to tool-support since profiles are supported by almost all UML tools. Therefore, in order to provide tool-support for their profile, all a profile developer has to do is to provide a reference implementation within existing framework of the chosen UML tool. So far as the UML version support is concerned, though at first glance one can see the balance between UML 1.x and UML 2.0, most recent approaches have already provided support for UML 2.0.

*Inclination to AspectJ:* Among approaches that are influenced from the feature set of an existing aspect-oriented programming language, AspectJ appears to be the most popular programming language. In the specific context of transforming aspect-oriented constructs from design to code, this inclination may reduce the model-system gap. However, where not applicable, determining the precise effects on this model-system relationship would require further investigation, which may be conducted through defining a mapping between constructs at different levels.

*Missing model- to- code mapping details:* Mapping of model constructs to those of programming language is an important concern that is greatly related to code generation from visual models. Analysis and design models may cover a lot of feature details on their own, yet their transformation to code may yield abundant deal of effort. It is evident from the results that there are only two approaches that have provided mapping of design models to code, among them there is only one approach (i.e. Ref[35]) that supports UML 2.

*Dearth of support for behavioral diagrams:* Almost all variations of structure diagrams have been used by AOM approaches except Klein [34], which provides support for a behavior diagram only. An inadequate support for behavioral modeling of crosscutting concerns is one of the key restraints on the practical use of AOM approaches in the model-driven engineering process.

*Lacking in traceability support:* Except for two approaches i.e. Jacobson [25] and Clarke [22], none of the approaches provides support for traceability with respect to all phases of software development. It is also a key issue to affect the suitability of an AOM approach to be integrated into MDE process.

*Insufficient scalability:* Scalability of approaches is another aspect in which AOM approaches need to improve. So far as the abstraction is concerned, almost all approaches support ways to hide the details of crosscutting behavior from modeler. However, the modeling examples which are presented in these studies, typically, do not exploit all features related to proving the scalability. While we have considered an approach as scalable if it can handle composition of simply three or more concerns, only six approaches could meet this criteria. However, among these six approaches Clarke [22], Jacobson [25], Cottenier [29], and J.Klein [35-37] have provided adequate proof of scalability of their approaches. Again, this is an area which is of high relevance in terms of applicability of AOM approaches to large-scale development and aspect-oriented code generation.

Table 1: Comparison of AOM approaches on basis of notation-specific criteria

| Serial | Approach | Extension mechanism | Programming lang. influence | Mapping availability | UML version support | UML diagrams support | Design process support | External traceability | Internal traceability | Scalability |
|---|---|---|---|---|---|---|---|---|---|---|
| 1 | Stein [18] | MM, UP | AJ | ✘ | 1.x | CD, CoD, SD, UC | ✘ | D->I | n/a | ✘ |
| 2 | Ho [19] | UP | n/a | ✘ | 1.x | CD | ✘ | D->I | ✓ | ✘ |
| 3 | Aldawud[20] | UP | n/a | ✘ | 1.x | CD, StD | ✓ | R->D | n/a | ✘ |
| 4 | Von [21] | MM | AJ | ✘ | 1.x | CD, SD, CoD | ✓ | R->D | ✓ | ✓ |
| 5 | Clarke [22] | MM | SOP, AJ, HJ | ✓ | 1.x | PD, CD, SD | ✓ | R->D->I | ✓ | ✓ |
| 6 | Jacobson [25] | MM | AJ, HJ | ✘ | 2.0 | CD, PD, UC, SD, AD, StD | ✓ | R->D->I | ✓ | ✓ |
| 7 | France [26] | UP, MM | n/a | ✘ | 2.0 | CD, PD, CmD | ✓ | A->D | ✓ | ✘ |
| 8 | Pawlak[27, 28] | UP, MM | AJ | ✘ | 1.x | CD | ✘ | D->I | ✘ | ✘ |
| 9 | Cottenier[29] | UP | n/a | ✘ | 2.0 | CD, DD, SD, StD | ✘ | ✘ | ✓ | ✓ |
| 10 | Fuentes [31] | UP | AJ | ✘ | 2.0 | CD, AD, StD | ✓ | ✘ | ✘ | ✓ |
| 11 | Katara[33] | UP | n/a | ✘ | 1.x | PD | ✘ | ✘ | ✘ | ✘ |
| 12 | Klein [34] | n/a | n/a | ✘ | 2.0 | SD | ✘ | ✘ | ✘ | ✘ |
| 13 | J.Klein[35-37] | MM | n/a | ✓ | 2.0 | CD, SD, StD | ✓ | ✘ | ✓ | ✓ |
| 14 | Whittle [38] | UP | n/a | ✘ | 2.0 | CD, AD, StD | ✓ | ✘ | ✘ | ✘ |

**Legend:**

| | | | | | |
|---|---|---|---|---|---|
| ✓ | Supported | I | Implementation | UC | Use case diagram |
| ✘ | Not supported | DD | Deployment diagram | MM | Meta-model extension |
| n/a | Not applicable | CD | Class diagram | StD | State diagram |
| AJ | AspectJ | CoD | Collaboration diagram | AD | Activity diagram |
| UP | UML profile | SD | Sequence diagram | CmD | Communication diagram |

| R | Requirements | SOP | SOP [46] | GP | General purpose |
|---|---|---|---|---|---|
| D | Design | HJ | Hyper/J | PD | Package diagram |

## 4.2. Composition

Table 2 summarizes the evaluation with respect to the *Composition* category.

*Prevalence of Element Asymmetry for representation of concerns:* In general, concerns are represented using e*lement asymmetry* (<<aspect>>), and the studied approaches possess an additional interesting pattern in this regard. In particular, when elements are asymmetric, all approaches apply the asymmetric composition mechanism (i.e. *composition asymmetry),* and they often use pointcut-advice mechanism for that. Further, rule asymmetry is followed to control pointcut-advice combinations. However, it should be noted here that this is not an essential characteristic, and that in principle, asymmetric composition may be employed to handle symmetry at element level.

*Moderate support for composition at modeling level:* The studied approaches often provide support for composition of concerns at modeling level, and among them half of the approaches provide both static and dynamic composition mechanisms. The composed model can be used for validation and simulation purposes, and it can also be used to generate code. However, the support for composition must be complemented with appropriate tool-support (which we will see later), or else the modelers will have to model each concern and then manually compose it.

*Popularity of UML as complete model:* UML seems to be the language of choice for approaches that provide composition of models, to represent the composed model. This inclination to UML seems quite natural and convincing since aspect-orientation can be seen as an extension to object-orientation, for which UML is the standard and most widely used modeling language. In this context, however, relying on UML has some obvious advantages. For example, for model transformation and/or code generation, one will be working on a source model which is defined using a standardized and well-defined meta-model. Typically, this ensures that the relationship between model and code will be based on standard meta-models, that is, the meta-model of UML and the meta-model of some target programming language like AspectJ.

Table 2: Comparison of AOM approaches on basis of composition-specific criteria

| Serial No. | Approach | Concern representation | Composition plan | Composed model | Concern interaction |
|---|---|---|---|---|---|
| 1 | Stein [18] | <<aspect>> Class | S | n/a | ✓ |
| 2 | Ho [19] | Stereotyped class | S | n/a | ✗ |
| 3 | Aldawud[20] | <<aspect>> Class in AOSD_Package, State machine regions | n/a | n/a | ✓ |
| 4 | Von [21] | Enhanced class | S | Extended UML | ✓ |
| 5 | Clarke [22] | <<theme>> Package | S | Themes | ✓ |
| 6 | Jacobson [25] | <<use case slice>> Package, <<aspect>> Classifier | n/a | n/a | ✓ |
| 7 | France [26] | Package diagram parameterized with templates | S | UML | ✓ |
| 8 | Pawlak[27, 28] | <<aspect>> Class | n/a | n/a | ✗ |
| 9 | Cottenier[29] | <<aspect>> Class | s, d | UML | ✓ |
| 10 | Fuentes [31] | <<aspect>> Class | s, d | UML | ✓ |
| 11 | Katara[33] | <<aspect>> Package | n/a | n/a | ✗ |
| 12 | Klein [34] | Sequence diagrams representing Pointcut and Advice | s, d | UML | ✗ |
| 13 | J.Klein[35-37] | Aspect package comprising of three parts: Class diagram, aspectual sequence diagram, aspectual statechart diagram | s, d | UML | ✓ |
| 14 | Whittle [38] | Scenarios – modeled using sequence diagrams with <<create>> and <<context>> stereotypes | s, d | UML | ✓ |

**Legend:**

| ✓ | Supported | s | Static | Asym | Asymmetric | n/a | Not applicable |
|---|---|---|---|---|---|---|---|
| ✗ | Not supported | d | Dynamic | Sym | Symmetric | | |

### 4.3. Maturity

Summary of comparison results from the perspective of *Maturity* of approaches are given in Table 3.

Table 3: Comparison of AOM approaches on basis of maturity-specific criteria

| Serial | Approach | Modeling example breadth | Modeling example depth | Real-world application | Information availability |
|--------|----------|--------------------------|------------------------|------------------------|--------------------------|
| 1 | Stein [18] | 2 | 2 | ✗ | 4 |
| 2 | Ho [19] | 1 | 2 | ✗ | 2 |
| 3 | Aldawud [20] | 1 | >2 | ✗ | 5 |
| 4 | Von [21] | 2 | 6 | ✗ | 3 |
| 5 | Clarke [22] | >5 | >15 | ✗ | >15 |
| 6 | Jacobson [25] | 1 | >3 | ✗ | 3 |
| 7 | France [26] | 3 | 1 | ✗ | >10 |
| 8 | Pawlak [27, 28] | 1 | 1 | ✓ | 3 |
| 9 | Cottenier [29] | 2 | >4 | ✓ | 9 |
| 10 | Fuentes [31] | >3 | >5 | ✗ | 6 |
| 11 | Katara [33] | 1 | >5 | ✗ | 2 |
| 12 | Klein [34] | 1 | 1 | ✗ | 9 |
| 13 | J.Klein[35-37] | 2 | >15 | ~ | 7 |
| 14 | Whittle [38] | 2 | >5 | ✗ | 4 |

**Legend:**

| ✓ | Supported | ✗ | Not supported | n | Number of examples, number modeled concerns, number of publications |
|---|-----------|---|---------------|---|---------------------------------------------------------------------|
| ~ | Partly supported | n/a | Not applicable | | |

*Moderate application of complicated example scenarios:* The use of complex modeling examples, which model three or more concerns, is moderately considered by existing approaches. Some of the studied approaches use trivial examples which are not sufficient to demonstrate and validate the strength of a modeling notation. However, Theme/UML (see Clarke [22]) and Reusable Aspect Models (see J.Klein[35-37]) have been demonstrated using nontrivial modeling scenarios comprising of over 15 concerns each. In fact, the latter has used two well-known case studies, which have been presented in literature specifically to determine the strengths and weaknesses of aspect-oriented modeling approaches. Similarly Cottenier et al. set a good example in this regard by validating their approach with the help of a complex real-world application for telecommunication industry.

 *Lack of real-world applications:* So far, very few AOM approaches have been tested in a real-world project i.e. the approaches of Pawlak et al. and Cottenier et al. In order to prove the suitability of AOM approaches for large-scale development, their application in real-world projects is a mandatory requirement.

## 4.4. Tool-support

Lack of *tool-support for composition:* A majority of approaches has provided tool-support for modeling only which is largely due to their use of UML profile mechanism. On one hand, the support for composition of models is an essential requirement for approaches which handle details of composition at modeling level only and generate object-oriented code, since without this support it would not be possible to obtain a composed view of the system. On the other hand, it is desirable for other approaches as well that defer composition to implementation level, since in this case the same can be used for activities such as simulation and validation of model. As shown in Table 4, half of the approaches have not discussed any mechanism for composition of models.

Table 4 summarizes the evaluation with respect to the *Tool-support* category.

*Lack of tool-support for composition:* A majority of approaches has provided tool-support for modeling only which is largely due to their use of UML profile mechanism. On one hand, the support for composition of models is an essential requirement for approaches which handle details of composition at modeling level only and generate object-oriented code, since without this support it would not be possible to obtain a composed view of the system. On the other hand, it is desirable for other approaches as well that defer composition to implementation level, since in this case the same can be used for activities such as simulation and validation of model. As shown in Table 4, half of the approaches have not discussed any mechanism for composition of models.

Table 4: Comparison of AOM approaches on basis of tool-support-criteria

| Serial No. | Approach | Modeling support | Composition support | Code generation support |
|---|---|---|---|---|
| 1 | Stein [18] | ✓ | ✗ | ✗ |
| 2 | Ho [19] | ✗ | ✗ | ✗ |
| 3 | Aldawud [20] | ✓ | ✗ | ✗ |
| 4 | Von [21] | ✗ | ✗ | ✗ |
| 5 | Clarke [22] | ✗ | ✗ | ✗ |
| 6 | Jacobson [25] | ✗ | ✗ | ✗ |
| 7 | France [26] | ✓ | ✓ | ✗ |
| 8 | Pawlak [27, 28] | ✓ | ✗ | ~ |
| 9 | Cottenier [29] | ✓ | ✓ | ✓ |
| 10 | Fuentes [31] | ✓ | ✓ | ✗ |
| 11 | Katara [33] | ✓ | ✓ | ✗ |
| 12 | Klein [34] | ✓ | ~ | ✗ |
| 13 | J.Klein[35-37] | ✓ | ✓ | ✗ |
| 14 | Whittle [38] | ✓ | ✓ | ✗ |

**Legend:**

| ✓ | Supported | ✗ | Not supported | ~ | Partly supported |
|---|---|---|---|---|---|

*Nonexistence of support for aspect-oriented code generation:* The investigation of tool-support for code generation reveals more interesting results. In this regard, the support provided by Cottenier et al. and Pawlak et al. has limitations from different perspectives. Specifically, Cottenier el al. generate object-oriented code only, even though they cover several aspects of code generation for the specific domain they considered (telecommunication applications). Pawlak et al. have addressed generation of aspect-oriented code but to a very limited extent, while they define only the mapping of their constructs to code in AspectJ language.

## 4.5. Principal findings in the context of aspect code generation

So far we have focused on comparing existing aspect-oriented modeling approaches for prospects along their use as a source notation in a model-driven environment. However, as previously given, we are particularly interested in investigating their potential to support enhancements for aspect-oriented code generation. This is because we believe that the automatic generation of aspect-oriented code can positively enhance the benefits of these modeling approaches. However, as we have mentioned in Section 4.1- Section 4.5, current modeling approaches lack from many perspectives if examined for the purpose of aspect code generation. In the following we sum up the important conclusions that can be drawn from results of comparison presented in previous subsections.

*Adoption of approaches requires elaboration of model-code relationship:* All modeling approaches provide fair support for abstracting the complex details related to handling of crosscutting functionality. However, in order to use them for the purpose of aspect-oriented code generation, the complexities involved in their transition to code need to be studied further. It is not often straightforward to map the concepts in a modeling notation to constructs of a programming language. That is why software developers often find it difficult to transform even excellently developed models into code. Therefore, unless the model-code gap is eliminated, expecting a practical integration of aspect orientation and model-driven engineering, which generates industry-standard code, would be unrealistic. Thus, more work is needed that defines details of transformation of constructs at design level to code level.

*Guidance on the use of asymmetric vs. symmetric approaches missing:* As briefly noted in Section 4.2, currently one can find a pattern with regard to applying symmetry or asymmetry at composition level. However, in principle, any type of composition i.e. symmetric or asymmetric, may handle any type of representation of element. But the current approaches do not answer the question when to use asymmetric or a symmetric approach sufficiently. This may have implications for code generation approaches. For instance, approaches that use element asymmetry will best suit the enhancement to support code generation targeted at programming languages that define aspects in an asymmetric way e.g. AspectJ. However, what will be the effect of a particular selection on the quality of design and/or produced code has not been explored by existing approaches yet.

*Tool-support needs integration with standard development environments:* Aspect-oriented techniques (and their integration with model-driven engineering to develop systems) are unlikely to be adopted in practice if not complemented with reliable tool support. The primary reason for this is the fact that manual weaving of aspect-oriented models is a fairly cumbersome and error-prone task. For interested readers, a detailed discussion of difficulties associated with behavior of a system in presence of aspects may be found in studies such as [47, 48]. This problem is further enlarged in the context of code generation, because of the raised complexity involved in handling of concerns at code level. Hence, further work is required to provide tool support for composition and code generation, and in this way to help increase the adoption of these approaches in practice.

*Strength of UML not fully exploited:* As evident from results in Table 1, majority of approaches support both structural as well as behavioral diagrams, thus allowing a modeler to consider both structure and behavior of a system. However, the selection of diagrams by different approaches is vastly different. UML diagrams possess varying levels of suitability for effectively modeling the dynamics of a particular system. Hence, it would be interesting to see how to support diagrams which are not supported currently. Moreover, only a few approaches provide composition of diagrams that they support for designing aspects. Therefore, it is imperative to go further in this context to benefit from full spectrum of UML diagrams.

*Missing guidelines on effectiveness of various modeling diagrams:* In principle, to model a system, a specific set of UML diagrams is chosen so that it can model a system in the most comprehensive and complete manner, from both structural as well as behavioral perspectives. This means that the final UML model must be representing the system in its entirety, with respect to its structure and behavior. This is particularly relevant when code is to be obtained from the model, since the completeness of output code would be directly linked with comprehensiveness of the input model. Current approaches provide no discussion on when to use one diagram or another, and what will be the impact of choosing a certain combination of diagrams. Hence, it would be valuable to investigate the relationship between various diagrams, the completeness of a model obtained by combining them and code obtained by transforming that design model into code.

## 5. In-depth comparison of Theme/UML and Reusable Aspect Models

The results of comparison of existing AOM approaches given previously show that the Reusable Aspect Models (RAM) and Theme/UML are two relatively mature approaches that can be used for the purpose of aspect-oriented code generation. In this section, we proceed to evaluate these approaches with respect to the details of code obtained from each approach. Specifically, we take a simple example application and model some of its relevant parts using both the RAM and Theme/UML approaches, and then we map these models to AspectJ code in order to carry out a thorough comparison of the obtained code.

### 5.1. The Online Book Store (OBS) System

As an appropriate modeling example to illustrate the RAM notation, an Online Book Store System has been adopted from model-driven engineering literature [49]. This system has previously been used in aspect-oriented software development literature to explain some AOM notations (e.g. [31]). It has to be emphasized here that in this section (and by selection of this example system), our intention is not to prove the strength of modeling notations, since where applicable, it has already been shown by means of the case studies mentioned in previous section. Instead, we have taken this example to apply the guidelines of both approaches with regards to modeling and mapping, and consequently, to get a basis for conducting a detailed analysis and comparison of both approaches.

In [49], the authors identify several use cases for the Online Book Store (OBS) System, for example, ordering of books, cancelling of order, approving of charge, delivering of order etc. However, mimicking the presentation of [31], we will focus on the ordering of books, which is described below:

(i) Ordering of books use case starts by the customer selecting a book and its desired quantity.

(ii) The customer adds more books if desired.

(iii) When the customer is finished with adding selected books to the order and wants to checkout, a message is sent to the credit card company to process the payment. This adds a few related requirements as well:

    a. Since all purchases in OBS System are made in euros whereas the credit card company carries out all operation in US dollars, currency conversion is required while exchanging messages with the credit card company.

    b. For privacy reasons, all communication with the credit card company must be encrypted.

(iv) If the payment is processed successfully, a shipping order is created, followed by delivering a message to the delivery company notifying that a new order is ready.

(v) In case an order is changed, the change is persisted. This adds another global requirement:

a.  All update operations for order must be persisted.

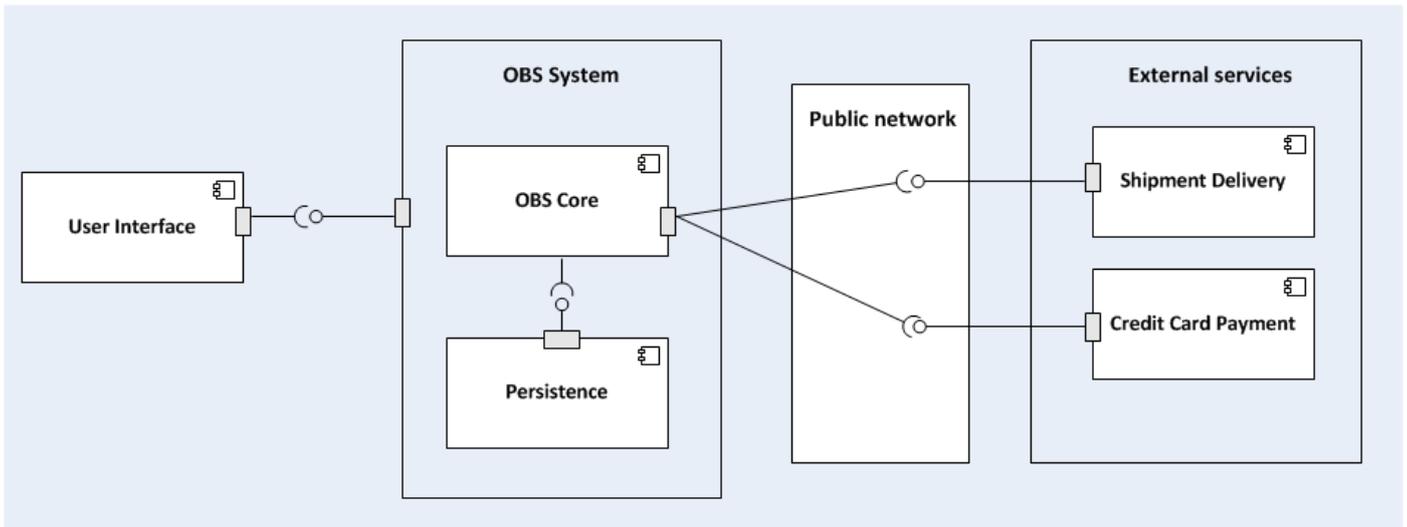

Figure 2: Excerpt of the high level architecture of OBS System

An excerpt of the high level architecture of the Online Book Store (OBS) System is shown in Figure 2. Apart from providing the core functionality of the OBS System, the OBS Core component is also responsible for handling the persistence of orders. The external services of payment and delivery are connected through a public network. The details of shipment delivery are uninteresting; therefore, we will concentrate only on payment, encryption and persistence of an order while performing functionality related to a new order. Figure 3 shows the high level communication of objects to handle a new order. The crosscutting functionality has been highlighted in grey regions. It should be observed that these aspectual parts will be woven by the weaver and not explicitly called. For example, a `chargeCreditCard` request dispatched to `a credit card payment controller` will result in inherently invoking `a currency converter`, which will convert euros to dollars and pass the order further, and `an encrypter`, which will encrypt the charge request to make it suitable for dispatching on public network. Therefore, if the aspectual part is eliminated, `an order` will directly be communicating with `a credit card payment controller`. This means that that the `chargeCreditCard` request is not actually sent to `CurrencyConverter`. Similarly, `paymentApproved` is not returned to `an encrypter`.

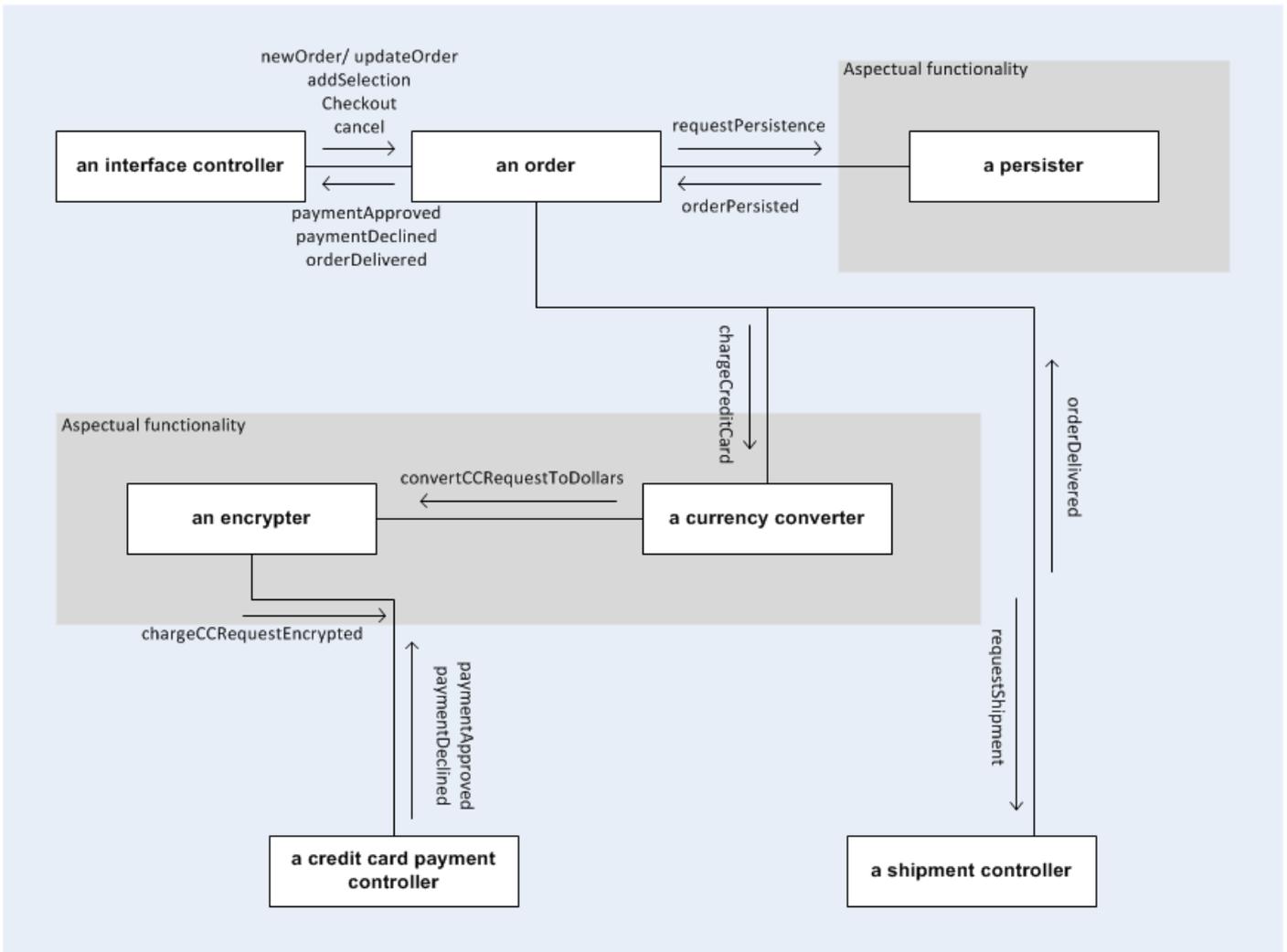

Figure 3: High level communication of objects to handle a new order

In the following, we model the crosscutting concerns of Online Book Store System using the selected modeling notations and proceed to mapping the models to AspectJ code.

### 5.2. The Reusable Aspect Models (RAM) approach

In this section, the RAM modeling approach is illustrated while presenting design of different aspects of the OBS System. This is followed by a brief description of the mapping process and the mapping of design to code.

#### 5.2.1. Design of Online Book Store System using RAM

As we briefly discussed in Section 3.1, the Reusable Aspect Models approach views aspects as concerns that are reused many times in an application, or across several applications. Therefore, this approach models any functionality that is reusable by means of an aspect. In OBS System, the crosscutting concerns of *Persistence*, *Currency Conversion*, and *Encryption* are modeled as separate reusable aspects. A brief description of these aspect models follows.

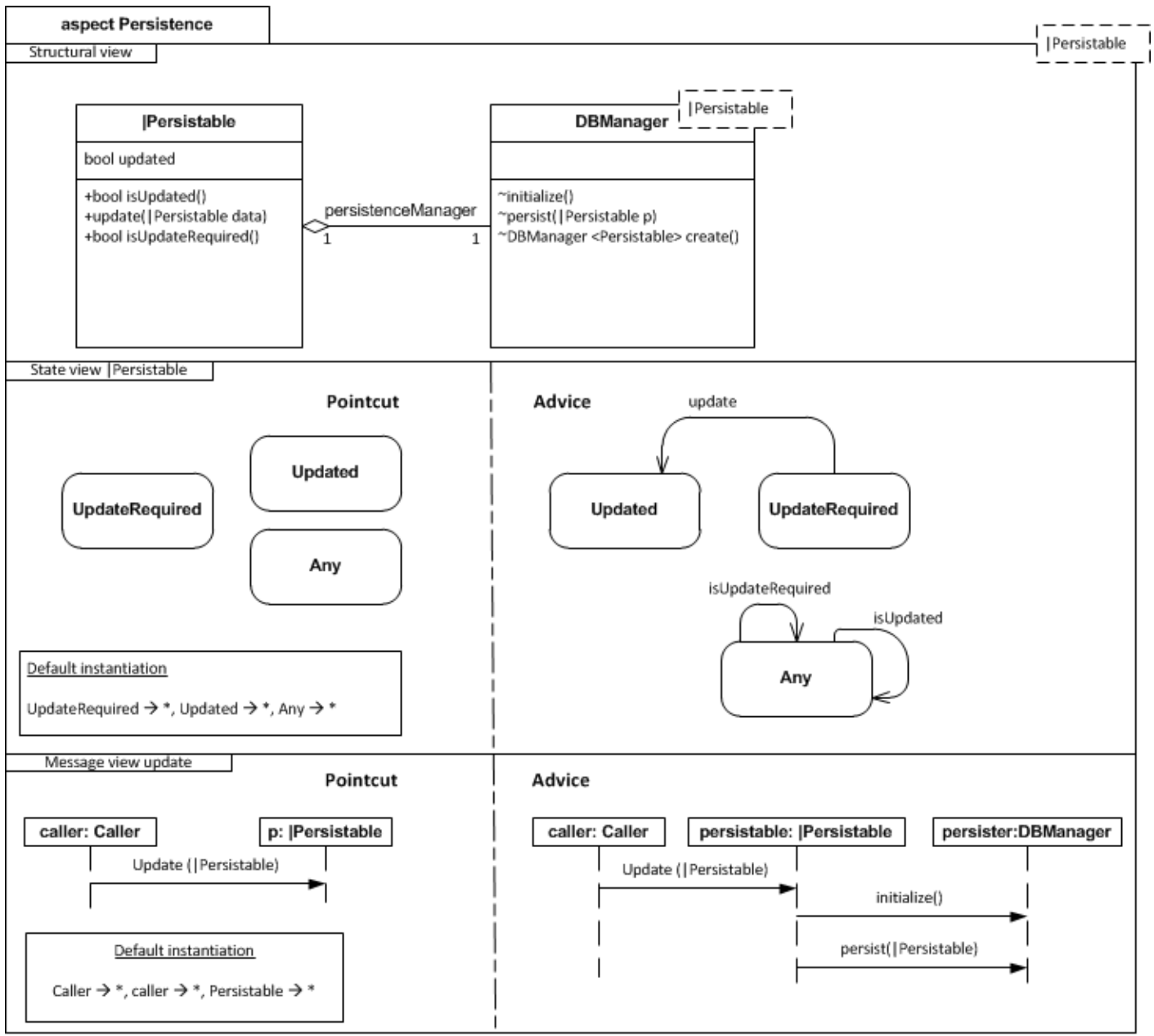

Figure 4: The Persistence aspect modeled using RAM

The *Persistence* aspect model for the OBS System is shown in Figure 4. The aspect is represented in the form of a special UML package in RAM. This aspect model comprises of different compartments. First compartment represents the structural view and is expressed using a UML class diagram. It is to be noted here that the classes in this compartment do not necessarily have to be complete. They may include methods and attributes that are relevant to the particular concern only. Such incomplete classes which are later to be composed with other classes by the weaver at time of instantiation of aspect or its binding to a base model, and methods whose names and signatures are to be determined later, are declared as *mandatory instantiation parameters*. The structural view of the *Persistence* aspect contains two classes, `DBManager` and `|Persistable`. The latter is an incomplete class and hence, in addition to having its name prepended with a "|" character, it is made prominent by specifying it as UML template parameters on top right corner of the structural view section. `|Persistence` provides methods to handle functions related to persistence. The public method

`isUpdateRequired` intends to identify situations where orders have really been updated and thus need persistence. In case an update operation has been performed, the persistence is carried out by `persist` method of `DBManager`, which is responsible for handling all the database-specific functionality related to persistence of objects.

As described previously, in RAM, following the structural compartment, several state view compartments are added, where each one corresponds to some class defined in the structural view of the aspect model. State view contains UML state diagrams to describe the internal states of the class that are relevant within the concern. For complete classes (i.e. standard classes) in the structural view, the state diagram is a standard state diagram. However, for incomplete classes, in which concerns are to be injected later, an *aspect state diagram* is defined which contains two parts: a pointcut and an advice. The pointcut part is used to define the states and transitions that are required in target state diagram, whereas the state diagram that replaces the occurrence of pointcut in the target state diagram is defined by the advice part. Similar to structural view, states are designated to be mandatory instantiation parameters in state view if their binding to states in standard state diagram does not exist. Such states are also annotated in a way similar to structural view. The pointcut view in the state view of Figure 4 shows three relevant states within the *Persistence* aspect i.e. `UpdateRequired`, `Updated` and `Any`. The advice part shows the capability of relevant states to call `update`, `isUpdateRequired` and `isUpdated` methods. The state view compartments in RAM are followed by message view compartments. The message view uses a UML sequence diagram to describe the sequence of messaging between different entities during the execution of a public method given in structural view of the aspect model. The message view compartment in the *Persistence* aspect model defines only one message view i.e. `update` message view. The pointcut in this message view states that all such situations are relevant in the context of this aspect where an `update` method is called on an instantiation of the |`Persistable` while some form of it is passed as an argument to this method. The advice states that all such method executions must also "persist" the updated order, by initializing a connection to the `DBManager` and calling its implementation of the `persist` method. It should be noted that prior to making a call to `persist` method, the method `initialize` is explicitly invoked in order to avoid repeated execution of functionality to establish and maintain database connectivity.

In order to instantiate and use an aspect, its mandatory instantiation parameters need to be mapped to concrete model elements in a target model. Therefore in this case, the class |`Persistable` shown as the UML template parameter on the top right corner of the *Persistence* aspect package must be mapped to some class in the target model.

Figure 5 shows the *CurrencyConversion* aspect modeled as a reusable aspect using the RAM notation. The structural view defines one incomplete class |`Convertible` to handle currency conversion. The `convert` method of this class is intended to be woven into all classes that need its functionality. In this regard, the specific details of exchange rates etc. as well as the details of conversion are to be handled by a complete class `Currency`. The *CurrencyConversion* aspect defines two state views, one for the incomplete class |`Convertible` and the other for complete class `Currency`. The |`Convertible` state view defines two relevant states `ConversionRequired` and `Converted`. The advice suggests that in order to accept calls to `convert`, an implementing object must be in `ConversionRequired` state. Since the |`Currency` state view defines the protocol for a standard class (and not an incomplete one), it takes the form of a standard state diagram.

In order to use the *CurrencyConversion* aspect, a modeler must map the class |`Convertible` shown as a mandatory instantiation parameter to one of the elements in target model.

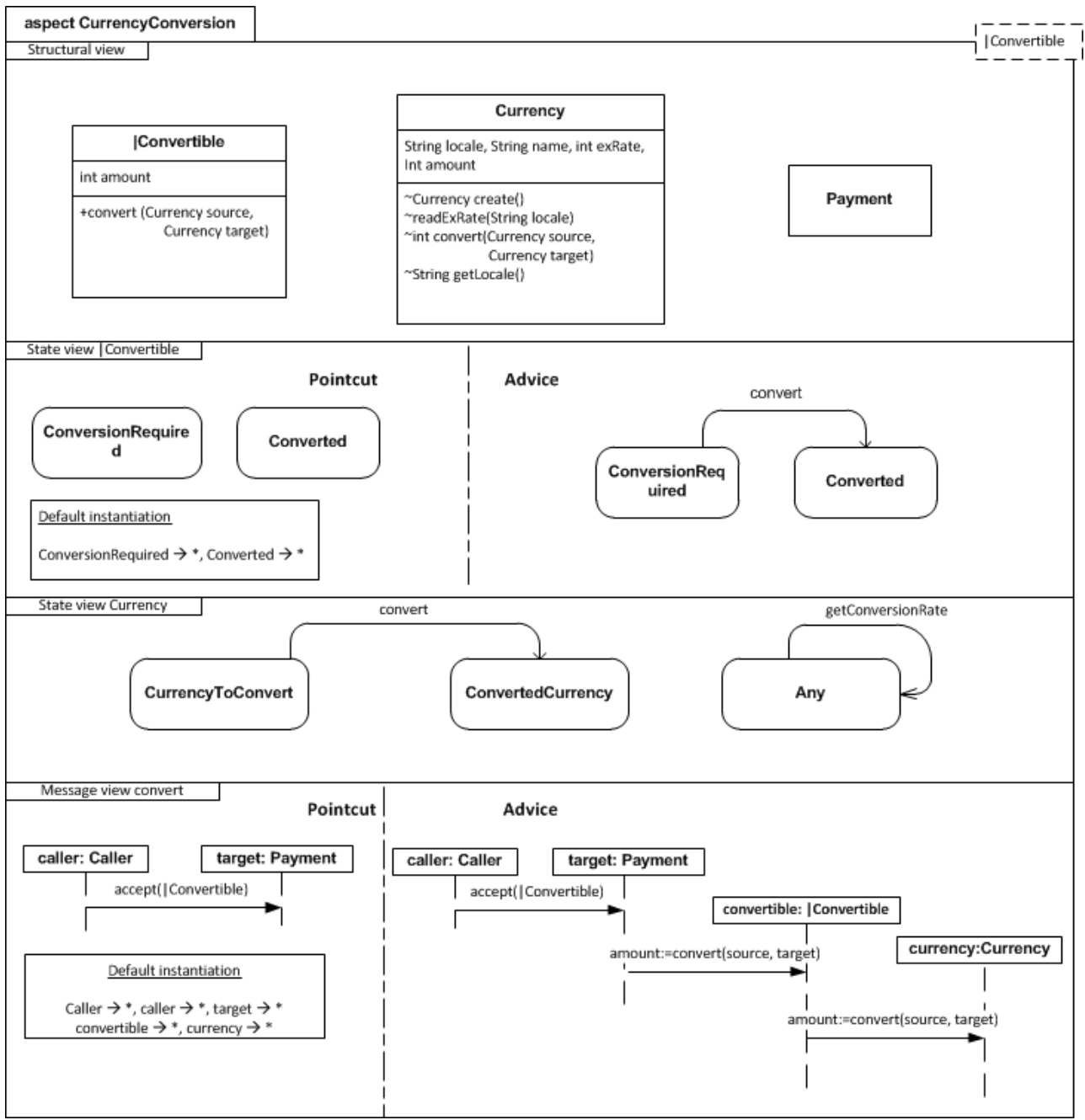

Figure 5: The Currency Conversion aspect modeled using RAM

Specifically, it defines three relevant states namely `CurrencyToConvert`, `ConvertedCurrency` and `Any`. First state is the one in which a call to `convert` method is possible, whereas the state diagrams shows that an implementing object may be in any state to make a call to `getConversionRate`. The message view compartment in the *CurrencyConversion* aspect model defines only one message view i.e. `convert` message view. In line with the description of the payment functionality given above, the pointcut in this message view takes all such calls to `accept` method as relevant wherein an instance of `|convertible` is passed as an argument. The advice specifies that in all such cases where payment is to be accepted, a conversion will need to be made by calling `convert` method while specifying appropriate currencies for the conversion as arguments.

This will follow a call to the `convert` method in `Currency` class that will handle the finer details of the required conversion.

Figure 6 shows the *Encryption* aspect which is a really simple aspect, in the sense that it contains only one class |Encryptable. Notice how it provides the functionality for encryption and decryption of an |Encryptable object, through its public methods `encrypt` and `decrypt`, respectively. Moreover, it allows to query the status of an object, telling if it was encrypted or not, through its public method `isEncrypted`. We consider the details of encryption and decryption uninteresting in the context of our study, thus none of the public methods in |Encryptable class involves object communication. As a result, the *Encryption* aspect contains no message view. The state view specifies that an instance of the |Encryptable class may have three states, here named `Encrypted`, `Decrypted` and `Any`. It further specifies that an instance must be in `Decrypted` state to be able to execute `encrypt`, and in `Encrypted` state to be able to call `decrypt` method. However, `isEncrypted` can be invoked by an object in `Any` state.

In a similar way to previously described aspects, the *Encryption* aspect may be instantiated and used by mapping the mandatory instantiation parameter i.e. the class |Encryptable to one of the elements in target model.

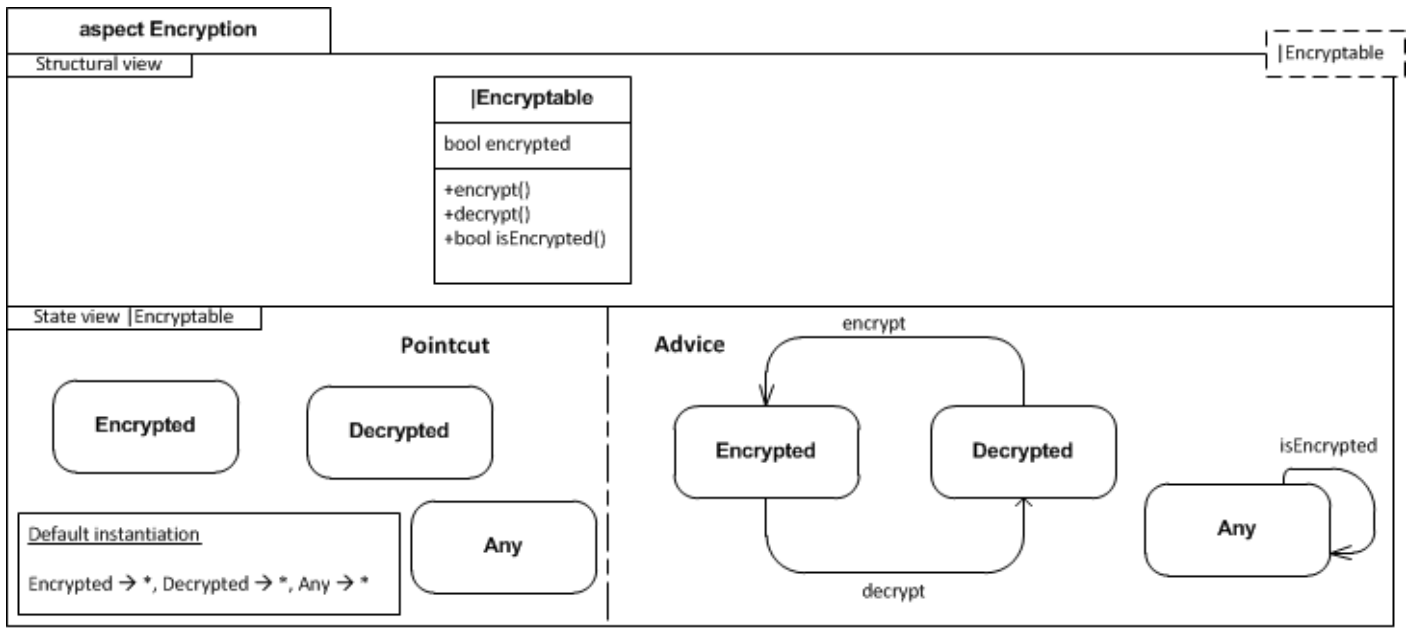

Figure 6: The Encryption aspect modeled using RAM

### 5.2.2.  Mapping of RAM design to AspectJ code

In [50], the authors propose a mapping scheme for mapping of RAM models to AspectJ code. In this section, we adopt their guidelines to the specific needs of this study, and present a mapping of aspects developed using RAM to AspectJ language.

In the following, we define the overall structure of the mapped code and some global items, and then we proceed to mapping different modeling elements along different views in aspects defined using RAM. Specifically, the global structure of the code is defined as follows:

A main package named `obss.ram` is created for the RAM project as a whole. This package will contain code for all aspects defined so far. Within this main package two sub packages named `obss.ram.aspects` and `obss.ram.conflictresolution` are created. In the proposed mapping technique, a sub package is created in the aspects package for each ordinary RAM aspect which has the same name as that of the aspect. This package contains all the artifacts of the RAM aspect. **Error! Not a valid bookmark self-reference.** shows three packages representing the three aspects modeled for OBS System.

```
1  obss.ram.aspects.persistence
2  obss.ram.aspects.encryption
3  obss.ram.aspects.currencyconversion
```

Figure 7: Ordinary aspect packages

Following the definition of sub packages that contain all artifacts of aspects, three required aspects are created directly into the main package. The three required aspects are shown in Figure 8.

```
1  obss.ram.AnnotationInheritance
2  obss.ram.ConfigurationEnforcement
3  obss.ram.AspectPrecedence
```

Figure 8: Required aspects in the main package

After defining the overall code architecture, we proceed to mapping different views belonging to individual aspects to AspectJ code, and start with the *Persistence* aspect. As described previously, the structural view of an aspect in RAM contains UML class diagrams and supports some additional features to specify methods or classes as mandatory instantiation parameters. While mapping classes in the structure view, first, the complete classes are mapped. The mapping approach of Kramer and Kienzle [50] propose a reuse mechanism to make use of the Java library, for classes and interfaces that resemble in structure or behavior to existing ones in Java. Since this is not the case for our modeled classes, we move on to generating their Java implementation from scratch. Therefore, for every complete class, we create a new public Java interface and an AspectJ aspect within the source file of this interface. In the next step, fields and methods are introduced into this interface using AspectJ's inter-type declaration mechanism. The code for `DBManager` complete class in the *Persistence* aspect is shown in Figure 9. Similarly, the mapping of `Currency` complete class for the *CurrencyConversion* aspect is shown in Figure 13.

```
public interface DBManager {
    //signatures of public methods in structural view
    void initialize();
    void persist (Persistable persistableObj);
}
aspect DBManagerAspect {
    //methods without message views
    public void DBManager.initialize() {
    }
    public void DBManager.persist(Persistable
                            persistableObj) {
    }
}
```

Figure 9: Mapping of the complete class `DBManager`

Apart from generating an interface, for complete classes in RAM aspects, a public class is created that implements that interface (generated in previous step). The name of this class is generated by appending the string "Impl" to the name of the modeled class (see Figure 10 and Figure 13). It should be noted that these classes are used for instantiation purpose only (as interfaces cannot be instantiated in Java) and are empty except for possible constructors.

```
public class DBManagerImpl {
    public DBManager() {
    }
}
```

Figure 10: Implementation class for `DBManager` interface

Following the complete classes, we map the incomplete classes in our aspects. The mapping of incomplete classes may be distinguished from that of the complete classes in that the process of matching to standard Java library is skipped and that an implementation class in not generated since the incomplete classes cannot be instantiated. Figure 11 shows the code obtained by mapping the incomplete class `Persistable`. Specifically, it shows that the attributes of |Persistable are mapped as variables of the respective types within the aspect using AspectJ's inter-type declarations. Associations are mapped as a special case of introducing Java fields. Therefore, a field is introduced whose name corresponds to the name of association in given aspect's structure view. Few rules that relate to multiplicity of the association are as follows:

- If the multiplicity is `0..1`, then the field is of the class type and is initialized to `0`.
- If the multiplicity is `1`, then the field will be of the same type but it will be initialized using a no-parameter constructor of the implementation class.
- If the multiplicity is `1..*`, then the type of corresponding field is `java.util.Set` and it is parameterized using the type of the associated class.

The `persistenceManager` field in `Persistable` interface represents an association of this class with `DBManager` (see Figure 11).

```
public interface Persistable {
      //public methods in structural view
      boolean isUpdated();
      void update(Persistable data);
      boolean isUpdateRequired();
}
aspect PersistableAspect {
      //attributes and associations in the structural view
      private boolean Persistable.updated = false;
      private DBManager Persistable.persistenceManager =
                        new DBManagerImpl();
      //methods without message views
      public boolean isUpdated() {
      }
      public boolean isUpdateRequired() {
      }
      // methods containing message views
      public void Persistable.update(Persistable data) {
            DBManager persister = new DBManager();
            persister.persist(data);
      }
}
```

Figure 11: Mapping of incomplete class `Persistable`

Code stubs are generated for methods that are not associated with a message view. There is one exception - methods which follow common patterns and make their implementation obvious e.g. getter, setter, and increment type methods, are fully generated in this step. However, this must be apparent from the method signatures. Like attributes, methods that have a message view are added using the inter-type declaration mechanism of AspectJ. The implementation of such methods is generated by adding method calls specified in the message view. See for example, the initialization of a `persister` object and its invocation of the `persist` method given in Figure 11, which is based on the message view of *Persistence* aspect in Figure 4.

```
@Target ({ElementType.TYPE})
@Retention (RetentionPolicy.RUNTIME)
public @interface PersistableClass {

}
```

Figure 12: Java annotation for mandatory instantiation parameter class `Persistable`

If an incomplete class is marked as a mandatory instantiation parameter, we create a Java annotation with the same name and the word "Class" appended to it. Figure 12 shows an annotation for the class `Persistable` marked as mandatory instantiation parameter in the *Persistence* aspect.

```
public interface Currency {
      //signatures of public methods in structural view
      void getConversionRate();
      void convert(Currency target);
}
aspect CurrencyAspect {
      //attributes
      String name; String locale; int exRate;

      //methods without message views
      protected void Currency.readExRate(String locale) {
            //details of currency database handling are
            //uninteresting, thus not represented in message view
      }
      public String getLocale() {
      }
      public int Currency.convert(Currency source,
                            Currency target {
      }
}

public class CurrencyImpl {
      public Currency() {
      }
}
```

<div align="center">Figure 13: Mapping of <code>Currency</code> complete class</div>

Following the same mapping definition, the `Convertible` interface in Figure 14 defines a method for the only public method in its structural view and provides its implementation within the aspect `ConvertibleAspect`. This is, however, limited to instantiating a new object of `Currency` and calling its `convert` method. The details of conversion were considered irrelevant to this study and hence omitted from the message view.

```
public interface Convertible {
      //signatures of public methods in structural view
      void convert(Currency source, Currency target);
}
aspect ConvertibleAspect {
      //methods with message views
      public void Convertible.convert(Currency source,
                             Currency target) {
          int amount = new Currency().convert(source, target);
      }
}
```

<div align="center">Figure 14: Mapping of the <code>Convertible</code>  incomplete class</div>

Figure 15 shows the code obtained for the Encryption aspect by applying the mapping rules. This is a rather simple aspect and its mapping is straightforward.

```
public interface Encryptable {
      //public methods in structural view
      void encrypt();
      void decrypt();
      boolean isEncrypted();
}
aspect EncryptableAspect {
      //attributes and associations in the structural view
      private boolean Encryptable.encrypted = false;
      //methods without message views
      public void encrypt() {
      }
      public void decrypt() {
      }
      public boolean isEncrypted() {
      }
}
```

Figure 15: Mapping of incomplete class `Encryptable`

### 5.3. The Theme/UML approach

In this section, the Theme/UML models of the selected aspects of the OBS System are presented. This is followed by a brief description of the mapping process, and the mapping of design to code.

#### 5.3.1. Design of OBS System using Theme/UML

As briefly described in Section 3, the Theme/UML approach makes a distinction between the "base" themes, and the "aspect" themes which refer to the crosscutting behavior. An aspect theme is differentiated from a base theme in that in addition to other behavior, it may define some behavior that is triggered by behavior in some other theme. As far as modeling process is concerned, first the triggered behavior needs to be identified and captured in the form of templates, and then the crosscutting behavior related to those templates is modeled [22]. Later, the base themes which are not affected by the crosscutting themes are modeled using the standard UML design process. Since in the scope of this study we are interested in design and mapping of aspects only, we proceed to modeling the crosscutting concerns identified for the Online Book Store System.

In Theme/UML, aspect themes are represented using a new complete unit of modularization similar to a package in standard UML with stereotype <<theme>>. This theme may comprise of any of the standard UML diagrams to model different views of the structure and behavior required for a concern to execute. The design of *Persistence* aspect model in Theme/UML notation is shown in Figure 16. As shown, the aspect theme design is similar to a standard UML package along with structural and behavioral diagrams, but is different in the way that it contains templates listed inside the theme package notation, and contains a sequence diagram for each of the templates grouping in the theme package.

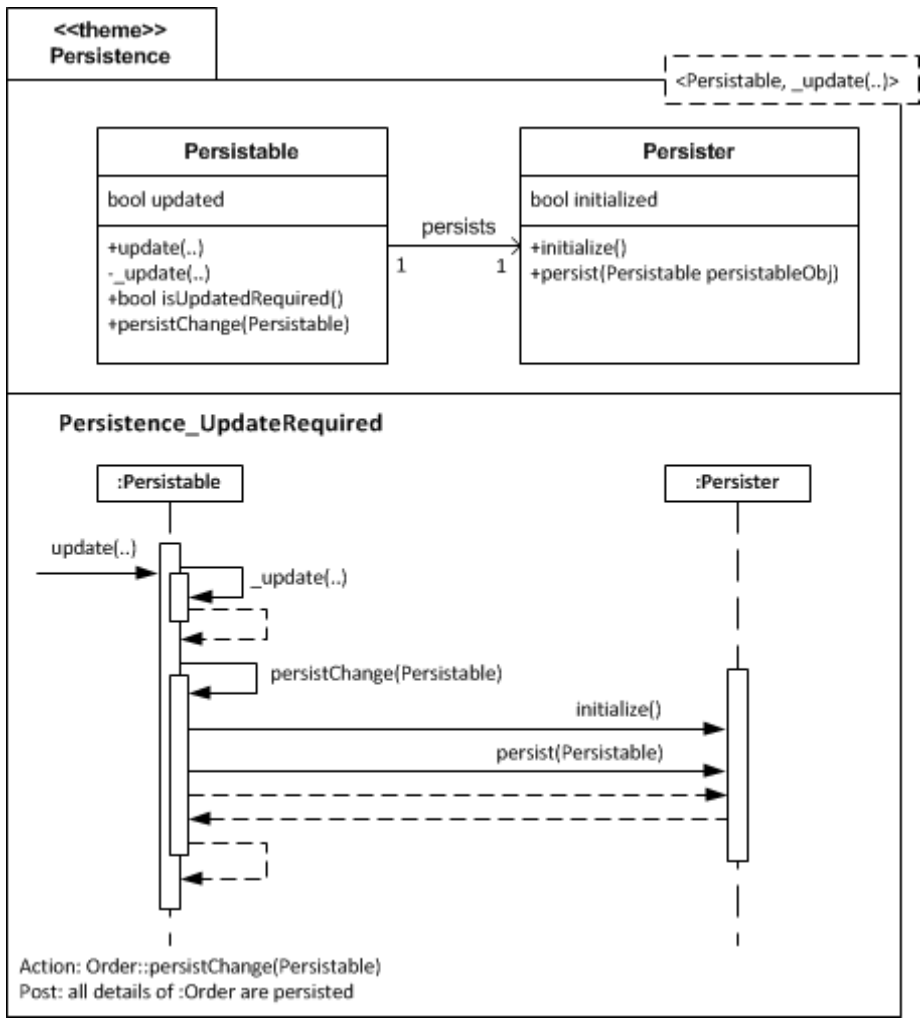

Figure 16: The Persistence aspect modeled using Theme/UML

It should be noted that each theme in this approach is intended to show only the classes and behavior that are necessary to represent its concepts. This means that each theme includes classes that it requires from its own perspective, and specifies them regardless of whether other themes also have classes to represent the same concepts. Any conflicts and overlaps that may arise are considered at a later stage. Therefore, initially, a design may contain different themes having different versions of the same class, each representing what is of interest for that specific theme. However, to crosscut a theme, the crosscutting theme requires abstract knowledge of the themes it crosscuts; crosscutting themes cannot operate independently. This external behavior in the crosscut theme is referred to by means of an extension of UML's notion of templates in a parameterized way; see for example the `update` operation of `Persistable` class exported as a template in Figure 16. The semantics of this declaration specify that the functionality of `update` method be augmented with some additional behavior defined by the sequence diagram given in the theme. Consequently, as given in the sequence diagram, a call to `update` method will essentially trigger execution of the `persistChange` method, which will in turn call `initialize` and `persist` functions on the `Persister` object.

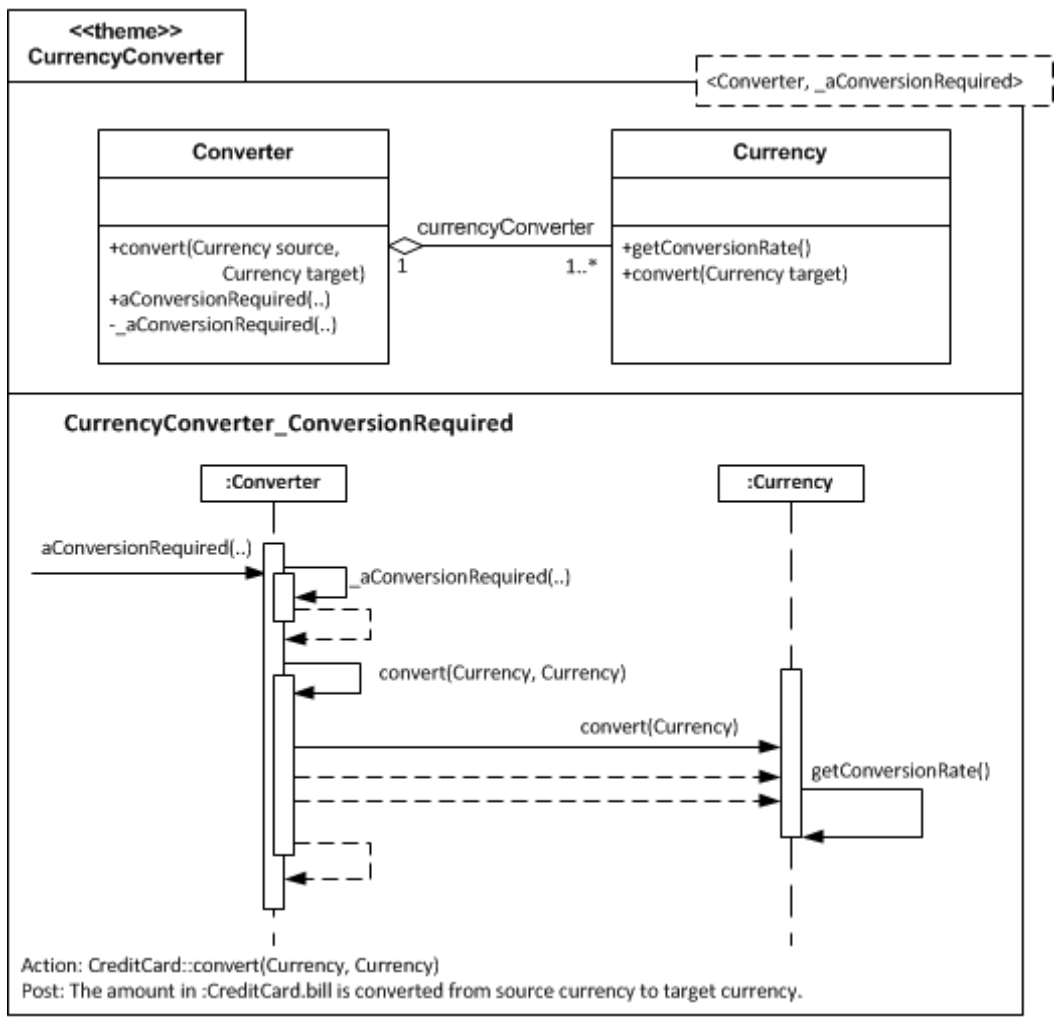

Figure 17: The Currency Conversion aspect modeled using Theme/UML

Next, the functionality related to currency conversion is modeled as a crosscutting theme named `CurrencyConverter` (See Figure 17). Specifically, this theme declares all such situations where a conversion is required as the crosscut points, and specifies the same by means of a template method `aConversionRequired` of the `Converter` class. Any invocation of this scenario results in supplementing the real behavior with behavior contained in `convert` method of the `Converter` class, which indeed invokes the `convert` function of `Currency` class and so on. The details of currency conversion have been intentionally omitted since we consider them uninteresting with regard to the discussion in this paper.

The theme/UML model of the *Encryption* aspect is not different from the RAM model of the same, mainly because it contains no behavioral logic. Therefore, we have not reproduced it in this section.

### 5.3.2. Mapping of Theme/UML design to code

Unlike the mapping approach for Reusable Aspect Models which starts with defining a global architecture of code, the Theme/UML approach directly maps the individual aspect themes. The Theme/UML mapping for the *Persistence* theme is shown in Figure 18. We start with defining an abstract aspect to represent the *Persistence* aspect. Within this aspect named `Persistence`, an interface is declared for the template class `Persistable`. Further, the mapping approach requires that an abstract pointcut be declared for each template method that initiates a sequence diagram. Therefore, we have declared the `update` pointcut that is associated with `after` advice relative to the sequence of behavior defined in the sequence diagram in Figure 16. Associations are defined using Java's native mechanism by defining variable of the target type (see e.g. `Persistable.updated`). Attributes and operations that are not declared as templates are implemented by defining an inter-type declaration on the class interface (see e.g. `Persistable.updated` and `Persistable.isUpdateRequired`). In end, classes without template operations are directly mapped as ordinary classes. This is in contrast to RAM's approach in which a class is mapped to an interface definition.

```
abstract aspect Persistence {
        //interfaces
        public interface Persistable { }
        //abstract attributes
        boolean Persistable.updated = false;
        Persister actualPersister = null;  //association
        //aspect methods
        public boolean Persistable.isUpdatedRequired() {
        }
        public void Persistable.persistChange(Persistable p) {
        }
        //abstract point cut
        public abstract pointcut update(Persistable persistableObject);
        //advice
        after (Persistable persistableObject) oncall:
                              update(persistableObject)
            {
                  actualPersister = new Persister();
                  actualPersister.initialize();
                  actualPersister.persist(persistableObject);
            }
}

// the Persister class
class Persister {
        private boolean initialized = false;
        public void initialize() {
        }
        public void persist(Persistable p) {
        }
}
```

Figure 18: Mapping of the Persistence Theme

Similarly, the same set of guidelines is applied to map the `CurrencyConversion` theme. The resultant code is shown in Figure 19. Notice how the one-to-many association named `currencyConverter` in Figure 17 has been mapped to a single reference of `Currency` class in the `CurrencyConverter` aspect. Theme/UML mapping approach does not give much detail on mapping different types of associations.

```
abstract aspect CurrencyConverter {
      //interfaces
      public interface Converter { }

      Currency curr = null;  //association: details of one-to-many
                             // association are absent
      //aspect methods
      public boolean Converter.convert() {
      }
      //abstract point cut
      public abstract pointcut aConversionRequired(Converter
                                                   objectToConvert);

      //advice
      after (Converter objectToConvert) oncall:
                         convert(objectToConvert)
          {
               curr = new Currency();
               curr.convert(objectToConvert);
          }
}
// the Currency class
class Currency {
      public void getConversionRate() {
      }
      public void convert(Currency target) {
      }
}
```

Figure 19: Mapping of the Currency Conversion Theme

### 5.4. Discussion

In Sections 5.2-5.3, we have presented the design of the Online Book Store System and its mapping to AspectJ code, using Reusable Aspect Models and Theme/UML approaches. It should be observed, however, that by means of using a trivial modeling example, the purpose was not to evaluate the strength of these approaches with regard to modeling. Rather, in line with the theme of this study, our intention was to apply their modeling and mapping techniques to obtain aspect-oriented code. In the following subsections we summarize the key findings of this activity and discuss some effects of these findings on the future research in this area. We divide our discussion into two parts: first we provide discussion from the modeling perspective and then we consider the code perspective.

#### 5.4.1.  The model perspective

As far as the models of Online Book Store System are concerned, there is a huge resemblance between the concepts supported by RAM and Theme/UML approaches. Specifically, both approaches use complete units of modularization i.e. aspects and themes respectively. Both approaches explicitly work to make their aspects as generic as possible through the use of UML templates with template parameters. In both approaches, the generic aspect model is instantiated by binding the aspect model's template parameters to elements of the target model. Nevertheless, in the process of applying guidelines provided by each approach to model our system, we have observed that the two approaches differ from some perspectives, and that in different cases one approach may have advantage over the other. These observations are briefly described in the following.

Unlike RAM, Theme/UML approach does not support the current version of UML. Apart from the observation that current UML tools often provide support for the current standard, there are few other differences in two standards, which may have some repercussions for the models developed using this approach. For instance, sequence diagrams are used as the main behavior modeling tool in Theme/UML, but several new features for UML 2.0 sequence diagrams such as *fragments*, *interaction occurrences*, *explicit representation of loops*, *notation for showing creation and destruction of objects* etc., are not supported [45]. Due to these features, in contrast with UML 1.x sequence diagrams, UML 2.0 sequence diagrams can work in two different forms: instance form and generic form. The instance form describes one possible interaction in a specific scenario, whereas the generic form documents all possible alternatives in a scenario.

In terms of relationship of aspect-oriented design models with overall software development life cycle, Theme/UML provides a detailed set of rules and guidance from analysis to implementation phases. In fact, a distinct part of theme approach i.e. Theme/Doc is used to handle analysis phase. In this regard, although RAM does not propose a complete methodology, but the existing large case studies provide sufficient insight into the design process, and guidelines to model systems effectively.

As far as the modeling diagrams are concerned, both approaches support structural as well as behavioral diagrams. Specifically, both approaches make use of class diagrams to model structure. For behavior modeling, RAM uses state diagrams and sequence diagrams. On the other hand, although theme approach is intended to support any type of diagrams, the support is not precisely defined except for sequence diagrams. Sequence diagrams are often suitable to showing collaborations among various objects involved in a single use case. However, they are not so good at precisely defining the behavior of an object [51]. The state diagrams and state machine specifications are considered the most effective and widely used method to specify behavior of a system [52, 53]. This way, the use of state diagrams in RAM approach is significant.

Despite the fact that the focus of this study is not on model weaving, some findings are interesting. We consider them worth mentioning because of the importance of model weaving for long-term research goals of model simulation, model testing and model debugging (cf. [30, 54, 55]). The aspects in RAM approach are woven using explicitly defined directives for the instantiation and binding. This explicit definition of the instantiation and binding directives improves internal traceability of models. No such directives are provided in Theme/UML models which use a *merge operator* to specify weaving of models. However, until now, the weaving with Theme/UML approach must be done manually.

As its name suggests, the reusability of aspect models is a major strength of the Reusable Aspect Models approach, whereas Theme/UML approach does not elaborate the reuse mechanism for its unit of modeling i.e. theme. In this regard, RAM supports the reuse of its aspect packages through creation of *aspect dependency chains*. An aspect providing complex functionality is modeled by decomposing it into many aspects that provide simpler functionality, and vice versa. This phenomenon also helps to hide the indirect dependencies of aspects from the user of an aspect.

Finally, tool support is another important factor while considering an aspect-oriented modeling approach. As we mentioned previously, unless supported by appropriate tools, an aspect-oriented modeling approach may raise the complexities of a modeling process. Theme/UML does not provide any tool support, neither for aspect modeling nor for aspect weaving. On the other hand, RAM comes with a tool developed in Kermeta [56] environment, which runs within the Eclipse Modeling Framework the use of Eclipse tools to edit, store and visualize models. The current RAM tool also supports the reusability of aspects and provides an inherent mechanism for consistency checks.

### 5.4.2. The code perspective

Reusable Aspect Models (RAM) and Theme/UML share some common points e.g. both are asymmetric approaches, both use a dedicated unit for encapsulation of aspects, both extend the UML meta model and define new constructs etc. As a result the code obtained from mapping to AspectJ is similar in many ways. However, few notes on the resultant code that may lead to selection of one approach over the other are presented in the following.

RAM approach explicitly considers the overall structure of code and provides few guiding principles to lead the mapping process [50]. Although no such information is available in current mapping approach of Theme/UML, a similar set of instructions may be applied, since the *aspect package* of RAM resembles the *theme package* of Theme/UML. With regard to overall structure, however, the role of required aspect packages in RAM (see Figure 8) is important. The `AnnotationInheritance` package is actually a repository of all declarations in an annotation hierarchy. As described previously, annotations are a way to map instantiation directives present in RAM models; these directives assign classes or methods to mandatory instantiation parameters (e.g. See Figure 12). As we discuss in the following, Theme/UML approach does not make use of annotations, hence no such structure mapping is available. The special AspectJ aspect named `ConfigurationEnforcement` is related to the reuse of aspects. Specifically, since RAM allows different configurations of an aspect at code level, each representing a variation of features, it needs some mechanism to keep a record of all possible variations. Thus for example, if two alternative variations of an aspect are not allowed to be used simultaneously, a definition of this configuration in the `ConfigurationEnforcement` aspect would specify that involved classes are not allowed to be marked with annotations corresponding to alternative variations. The `AspectPrecedence` aspect is also related to reuse hierarchies, and aspect dependency chains described in Section 5.4.1. Specifically, it keeps a list of all aspects that are reused in a context in the order of their reuse, to keep track of the precedence rules.

In Theme/UML ordinary classes i.e. complete classes are mapped to corresponding classes in Java containing plain variable declarations and function stubs. The approach of RAM to map complete classes to Java interfaces and a corresponding interface-implementing class is interesting. In the first step, to allow reuse of Java library, complete classes are checked to find their equivalent in a custom-built library of supported classes and interfaces from Java standard library. In case an equivalent is not located, it is mapped to a public Java interface that contains the aspect declaration. By doing so, instead of mapping a class to a standard class in Java, the approach allows merging of this class with other classes based on binding directives. Since *multiple inheritance* is not supported by Java, merging of a class with other classes can only be made possible by the means of its support for implementing multiple interfaces. Theme/UML approach does not consider reuse of existing Java classes or interfaces, and relies only on its own mechanism to generate classes from scratch.

As far as incomplete classes are concerned, both approaches implement them with interfaces. In Theme/UML, an interface implementing an incomplete class (or more specifically a pattern class in Theme approach) contains all non-template operations as methods. A template operation is mapped to an abstract method (if it is not associated with a sequence diagram) or an abstract pointcut (if it has a sequence diagram). On the other hand, RAM uses the Java annotations for mandatory instantiation parameters. In our opinion, although the code resultant from Theme/UML looks concise, mapping template operations in Theme/UML's way has some drawbacks. First, mapping a template operation to abstract method means that it needs to be bound to the target class by providing a corresponding implementation of the method with a delegating call. This is obviously more verbose than RAM's approach of marking existing methods with annotations. Second, mapping a template operation (with supplementary behavior) to an abstract point means that it requires the definition of concrete pointcuts for methods to be bound to target classes. Definition of concrete pointcuts will prevent this code from being used in other Java projects since pointcuts are not supported in pure Java. On the other hand, since annotations are supported by pure Java, the RAM's implementation can be reused in other standard Java projects.

Regarding the mapping of sequence diagrams, since Theme/UML is based on UML 1.x, the mapping for an important feature of UML 2 sequence diagrams i.e. notion of *Sequence Fragments* [45](sometimes referred to as *Interaction Frames* cf. [51]) could not be provided yet. Sequence fragments are essentially a control flow construct that is represented by a rectangular box surrounding a portion of sequence diagram and overlapping the section in which a fragment occurs. There are several different types of sequence fragments; currently only a few of them are supported by RAM approach. The supported fragments are: (1) option combination fragment, which is mapped to an *if* statement in Java, (2) alternation combination fragments, which are implemented using *if* followed by *else-if* statements in Java, (3) loop combination fragments, which are implemented using Java's *for* and *while* loops. There are several other types of fragments which are not supported by RAM approach i.e. *ref*, *assert*, *break*, *neg*, *par*, *region*, *sd* fragments.

Having considered current support for feature set of sequence diagrams, it has to be emphasized again that the sequence diagrams are not considered the best tool to precisely defining the behavior of an object. In this regard, as we mentioned previously, state diagrams are considered the most effective and widely used method of specifying system behavior [51, 57]. Therefore, they can be considered a more effective tool for full code generation than any other UML diagram. That is why, a considerable volume of literature (cf. [52]) is devoted to studying the implementation of state diagram in different programming languages. Currently, both RAM and Theme/UML have not considered implementation of state diagrams.

## 6. Conclusion

In this paper we provide a comparison of different AOM approaches with respect to their potential to integrate into an MDE environment by means of aspect-oriented code generation. For this task, we conducted the current study by using a well-defined comparison approach that has allowed us to identify and compare a set of 14 well-published AOM approaches in first part of this paper. In the second part, we have conducted a detailed comparison of 2 most comprehensive and most widely used approaches, Theme/UML and Reusable Aspect Models.

The results of first part of our study show that in order to achieve a true integration of aspect orientation and MDE, AOM approaches need to be improved from many perspectives. In particular, we have found that mostly AOM approaches suffer from problems including: (1) inadequate support for behavioral diagrams, (2) lack of scalability, (3) moderate consideration of complex application scenarios, (4) lack of application in the real-world, (5) lack of tool-support for composition, (6) absence of details of mapping from design to code, and (7) unavailability of tool-support for aspect code generation.

As regards the second part, the results show that Reusable Aspect Models and Theme/UML resemble in many ways with respect to handling modeling and mapping details of aspects. However, there are some points where RAM approach has advantage over Theme/UML such as: (1) its support for the latest UML version, (2) use of state diagrams for modeling behavior, (3) explicit definition of directives for the instantiation and binding, (4) its strength in terms of reusability of aspects both at design and code levels, (5) tool support, (6) improved handling of code structure, and (7) mapping of advanced features of sequence diagrams.

**References**

[1] A. Rashid, A. Moreira, J. Araujo, P. Clements, E. Baniassad, B. Tekinerdogan, Early aspects: Aspect-oriented requirements engineering and architecture design, in, Electronic Document. http://www.early-aspects.net/, 2006.

[2] T. Elrad, O. Aldawud, A. Bader, Aspect-Oriented Modeling: Bridging the Gap between Implementation and Design Generative Programming and Component Engineering, in: D. Batory, C. Consel, W. Taha (Eds.), Springer Berlin / Heidelberg, 2002, pp. 189-201.

[3] G. Kiczales, J. Lamping, A. Mendhekar, C. Maeda, C. Lopes, J.-M. Loingtier, J. Irwin, Aspect-oriented programming, in: M. Aksit, S. Matsuoka (Eds.) ECOOP'97 — Object-Oriented Programming, Springer Berlin / Heidelberg, 1997, pp. 220-242.

[4] W. Harrison, H. Ossher, P. Tarr, Asymmetrically vs. symmetrically organized paradigms for software composition, in, 2002.

[5] J. Hannemann, G. Kiczales, Design pattern implementation in Java and aspectJ, SIGPLAN Not., 37 (2002) 161-173.

[6] A. Garcia, C. Sant'Anna, E. Figueiredo, U. Kulesza, C. Lucena, A.v. Staa, Modularizing design patterns with aspects: a quantitative study, in: Proceedings of the 4th international conference on Aspect-oriented software development, ACM, Chicago, Illinois, 2005, pp. 3-14.

[7] L. Fuentes, P. Sánchez, Execution of Aspect Oriented UML Models, in: D. Akehurst, R. Vogel, R. Paige (Eds.) Model Driven Architecture- Foundations and Applications, Springer Berlin / Heidelberg, 2007, pp. 83-98.

[8] N. Cacho, C. Sant'Anna, E. Figueiredo, A. Garcia, T. Batista, C. Lucena, Composing design patterns: a scalability study of aspect-oriented programming, in: Proceedings of the 5th international conference on Aspect-oriented software development, ACM, Bonn, Germany, 2006, pp. 109-121.

[9] A. Hovsepyan, R. Scandariato, S.V. Baelen, Y. Berbers, W. Joosen, From aspect-oriented models to aspect-oriented code?: the maintenance perspective, in: Proceedings of the 9th International Conference on Aspect-Oriented Software Development, ACM, Rennes and Saint-Malo, France, 2010, pp. 85-96.

[10] A. Mehmood, D.N.A. Jawawi, Aspect-oriented model-driven code generation: A systematic mapping study, Information and Software Technology, 55 (2013) 395-411.

[11] M. Wimmer, A. Schauerhuber, G. Kappel, W. Retschitzegger, W. Schwinger, E. Kapsammer, A survey on UML-based aspect-oriented design modeling, ACM Comput. Surv., 43 (2011) 1-33.

[12] R. Chitchyan, A. Rashid, P. Sawyer, A. Garcia, M.P. Alarcon, J. Bakker, B. Tekinerdogan, S. Clarke, A. Jackson, Survey of Aspect-Oriented Analysis and Design Approaches. Technical Report AOSD. Europe Deliverable D11, AOSD-Europe-ULANC-9, in, Lancaster University, May 2005.

[13] A.M. Reina, J. Torres, M. Toro, Separating concerns by means of UML-profiles and metamodels in PIMs, in: O. Aldawud, G. Booch, J. Gray, J.o. Kienzle, D. Stein, Mohamed, F. Akkawi, T. Elrad (Eds.) The 5th Aspect-Oriented Modeling Workshop In Conjunction with UML 2004, 2004.

[14] S.O.d. beeck, E. Truyen, N. Bouck'e, F. Sanen, M. Bynens, W. Joosen, A Study of Aspect-Oriented Design Approaches, in, Department of Computer Science K.U. Leuven, 2006.

[15] A. Mehmood, D.N.A. Jawawi, A comparative survey of aspect-oriented code generation approaches, in: Software Engineering (MySEC), 2011 5th Malaysian Conference in, 2011, pp. 147-152.

[16] J. Stanley M. Sutton, I. Rouvellou, Concern modeling for aspect-oriented software development, in: R. Filman, T. Elrad, S. Clarke, M. Aksit (Eds.) Aspect-Oriented Software Development, Addison-Wesley, 2005, pp. 479–505.

[17] D. Suvee, W. Vanderperren, D. Wagelaar, V. Jonckers, There are no Aspects, Electron. Notes Theor. Comput. Sci., 114 (2005) 153-174.

[18] D. Stein, S. Hanenberg, R. Unland, An UML-based aspect-oriented design notation for AspectJ, in: Proceedings of the 1st international conference on Aspect-oriented software development, ACM, Enschede, The Netherlands, 2002, pp. 106-112.

[19] W.-M. Ho, J.-M. Jezequel, F. Pennaneac'h, N. Plouzeau, A toolkit for weaving aspect oriented UML designs, in: Proceedings of the 1st international conference on Aspect-oriented software development, ACM, Enschede, The Netherlands, 2002, pp. 99-105.

[20] T. Aldawud, Bader, A.,Tzilla Elrad, UML profile for aspect-oriented software development, in: The Third International Workshop on Aspect Oriented Modeling., 2003.

[21] V.F.G. Chavez, A model-driven approach for aspect-oriented design, in, Pontif´ıcia Universidade Cat´olica do Rio de Janeiro, 2004.

[22] S. Clarke, E. Baniassad, Aspect-Oriented Analysis and Design: The Theme Approach, Addison Wesley Object Technology, 2005.

[23] E. Baniassad, S. Clarke, Theme: an approach for aspect-oriented analysis and design, in: Software Engineering, 2004. ICSE 2004. Proceedings. 26th International Conference on, 2004, pp. 158-167.

[24] S. Clarke, R.J. Walker, Towards a standard design language for AOSD, in: Proceedings of the 1st international conference on Aspect-oriented software development, ACM, Enschede, The Netherlands, 2002, pp. 113-119.

[25] I. Jacobson, P.-W. Ng, Aspect-Oriented Software Development with Use Cases, Addison-Wesley Professional, 2004.

[26] R. France, I. Ray, G. Georg, S. Ghosh, Aspect-oriented approach to early design modelling, Software, IEE Proceedings -, 151 (2004) 173-185.

[27] R. Pawlak, L. Seinturier, L. Duchien, L. Martelli, F. Legond-Aubry, G. Florin, Aspect-Oriented Software Development with Java Aspect Components, in: R. Filman, T. Elrad, S. Clarke, M. Aksit (Eds.) Aspect-oriented software development, Addison-Wesley, 2005, pp. 343-369.

[28] R. Pawlak, L. Duchien, G. Florin, F. Legond-Aubry, L. Seinturier, L. Martelli, A UML Notation for Aspect-Oriented Software Design, in: AO modeling with UML workshop at the AOSD'02., 2002.

[29] T. Cottenier, A.v.d. Berg, T. Elrad, Joinpoint Inference from Behavioral Specification to Implementation, in: E. Ernst (Ed.) ECOOP 2007 – Object-Oriented Programming, Springer Berlin / Heidelberg, 2007, pp. 476-500.

[30] T. Cottenier, A.v.d. Berg, T. Elrad, The Motorola WEAVR: Model Weaving in a Large Industrial Context, in: Proceedings of the 6th International Conference on Aspect-oriented Software Development (AOSD '07), 2007.

[31] L. Fuentes, P. Sanchez, Designing and Weaving Aspect-Oriented Executable UML models, Journal of Object Technology, 6 (2007) 109-136.

[32] L. Fuentes, P. Sanchez, Towards executable aspect-oriented UML models, in: Proceedings of the 10th international workshop on Aspect-oriented modeling, ACM, Vancouver, Canada, 2007, pp. 28-34.

[33] M. Katara, S. Katz, A concern architecture view for aspect-oriented software design, Software and Systems Modeling, 6 (2007) 247-265.

[34] J. Klein, F. Fleurey, J.-M. Jézéquel, Weaving Multiple Aspects in Sequence Diagrams, in: A. Rashid, M. Aksit (Eds.) Transactions on Aspect-Oriented Software Development III, Springer Berlin / Heidelberg, 2007, pp. 167-199.

[35] J. Klein, J. Kienzle, Reusable Aspect Models, in: 11th Workshop on Aspect-Oriented Modeling, Nashville, TN, USA, 2007.

[36] J. Kienzle, W.A. Abed, J. Klein, Aspect-oriented multi-view modeling, in: Proceedings of the 8th ACM international conference on Aspect-oriented software development, ACM, Charlottesville, Virginia, USA, 2009, pp. 87-98.

[37] J. Kienzle, W. Al Abed, F. Fleurey, J.-M. Jézéquel, J. Klein, Aspect-Oriented Design with Reusable Aspect Models, in: S. Katz, M. Mezini, J. Kienzle (Eds.) Transactions on Aspect-Oriented Software Development VII, Springer Berlin / Heidelberg, 2010, pp. 272-320.

[38] J. Whittle, A. Moreira, J. Arajo, P. Jayaraman, A. Elkhodary, R. Rabbi, An Expressive Aspect Composition Language for UML State Diagrams, in: G. Engels, B. Opdyke, D. Schmidt, F. Weil (Eds.) MoDELS, Springer, 2007, pp. 514-528.

[39] J. Whittle, P. Jayaraman, MATA: A Tool for Aspect-Oriented Modeling based on Graph Transformation, in: 11th AOM Workshop, 2008.

[40] J. Whittle, P. Jayaraman, A. Elkhodary, A. Moreira, J. Araújo, MATA: A Unified Approach for Composing UML Aspect Models Based on Graph Transformation, in: S. Katz, H. Ossher, R. France, J.-M. Jézéquel (Eds.) Transactions on Aspect-Oriented Software Development VI, Springer Berlin / Heidelberg, 2009, pp. 191-237.

[41] S. Clarke, R.J. Walker, Composition patterns: an approach to designing reusable aspects, in: Proceedings of the 23rd International Conference on Software Engineering, IEEE Computer Society, Toronto, Ontario, Canada, 2001, pp. 5-14.

[42] F. Fleurey, B. Baudry, R. France, S. Ghosh, A Generic Approach For Automatic Model Composition, in: 11th AOM Workshop, 2008.

[43] L. Fuentes, P. Sánchez, Dynamic Weaving of Aspect-Oriented Executable UML Models, in: S. Katz, H. Ossher, R. France, J.-M. Jézéquel (Eds.) Transactions on Aspect-Oriented Software Development VI, Springer Berlin / Heidelberg, 2009, pp. 1-38.

[44] P. Desfray, UML Profiles Versus Metamodeling Extensions: An Ongoing Debate, in: Workshop on UML in the.COM Enterprise: Modeling Corba Components, XML/XMI and Metadata, Palm Springs, CA, USA, 2000.

[45] OMG, Unified Modelling Language Specification: Superstructure v2.2 in, http://www.omg.org/spec/UML/2.2/, 2009.

[46] S. Clarke, W. Harrison, H. Ossher, P. Tarr, Subject-oriented design: towards improved alignment of requirements, design, and code, SIGPLAN Not., 34 (1999) 325-339.

[47] C. Clifton, G. Leavens, A Design Discipline and Language Features for Formal Modular Reasoning in Aspect-Oriented Programs, in: Technical Report 05-23, 2005.

[48] G. Kiczales, M. Mezini, Aspect-oriented programming and modular reasoning, in: Proceedings of the 27th international conference on Software engineering, ACM, St. Louis, MO, USA, 2005, pp. 49-58.

[49] S.J. Mellor, M. Balcer, Executable UML: A Foundation for Model-Driven Architectures, Addison-Wesley Longman Publishing Co., Inc., 2002.

[50] M. Kramer, J. Kienzle, Mapping Aspect-Oriented Models to Aspect-Oriented Code, in: J. Dingel, A. Solberg (Eds.) Models in Software Engineering, Springer Berlin / Heidelberg, 2011, pp. 125-139.

[51] M. Fowler, UML Distilled: A Brief Guide to the Standard Object Modeling Language (3rd Edition), {Addison-Wesley Professional}, 2005.

[52] E. Dom´nguez, B. Pérez, Á.L. Rubio, M.a.A. Zapata, A systematic review of code generation proposals from state machine specifications, Information and Software Technology, 54 (2012) 1045-1066.

[53] I.A. Niaz, Automatic Code Generation From UML Class and Statechart Diagrams, in: Graduate School of Systems and Information Engineering., University of Tsukuba, Ph.D. Thesis., 2005.

[54] E. Long, A. Misra, J. Sztipanovits, Increasing productivity at Saturn, Computer, 31 (1998) 35-43.

[55] P. Baker, S. Loh, F. Weil, Model-Driven Engineering in a Large Industrial Context — Motorola Case Study, in: L. Briand, C. Williams (Eds.) Model Driven Engineering Languages and Systems, Springer Berlin / Heidelberg, 2005, pp. 476-491.

[56] P.-A. Muller, F. Fleurey, J.-M. J, #233, quel, Weaving executability into object-oriented meta-languages, in: Proceedings of the 8th international conference on Model Driven Engineering Languages and Systems, Springer-Verlag, Montego Bay, Jamaica, 2005, pp. 264-278.

[57] P. Rittgen, Enterprise Modeling and Computing With UML, {IGI Global}, 2006.